\documentstyle[12pt]{article}
\begin{document}
\begin{titlepage}
\begin{flushright}{FAU-TP3-98/4}
\end{flushright}
\vskip 3.0cm
\begin{center}
{\Large {\bf Polyakov Loop Dynamics in the \\ Center Symmetric Phase}}
\vskip 0.5cm
F. Lenz and M. Thies
\vskip 0.2cm
{\it Institute for Theoretical Physics III, University of
Erlangen-N\"urnberg, \\
 Staudtstr. 7, 91058 Erlangen, Germany}
\end{center}
\vskip 3.0cm
\begin{abstract}
A study of the center symmetric phase of  SU(2)
Yang Mills theory is presented. Realization of the
center symmetry is shown to result from non-perturbative gauge fixing.
Dictated by the center symmetry, this phase exhibits already at the
perturbative level confinement like properties. The analysis is performed
by investigating the dynamics of the Polyakov loops. The ultralocality of
these degrees of freedom implies significant changes in the vacuum structure
of the theory. General properties of the confined phase and of the
transition to the deconfined phase are discussed. Perturbation theory built
upon the vacuum of ultralocal Polyakov loops is presented and used
to calculate, via the Polyakov loop
correlator, the static quark-antiquark potential.
\end{abstract}

\end{titlepage}

\section{Introduction}
Confinement of the elementary degrees of freedom is a fundamental
property of
Quantum Chromodynamics (QCD). It  has been subject of many investigations,
and a
variety of mechanisms have been proposed for its explanation. Here
we mention as particularly relevant for our present
work the restriction in the range of the functional integration due
to the presence of
Gribov horizons in gauge fixed formulations \cite{GRIB78}, or the
appearance and possible condensation
of magnetic monopoles in Abelian projected descriptions \cite{tH81}.
Despite considerable analytical
and numerical efforts and largely due to the gauge dependence of
most of the
discussed mechanisms, no definite picture has yet emerged nor have
interrelations between
these mechanisms been established.

The focus of our study of the confining phase of QCD will be on the
center symmetry \cite{Suss79,McLerr81,Kuti81}
and the associated order-fields,
the Polyakov loop
variables \cite{Poly78,tHooft79,Svetitsky}.
Irrespective of the details of the dynamics which give rise to
confinement, this symmetry
must be realized in the confining and spontaneously broken
in the ``quark-gluon plasma''
phase. Unlike in lattice gauge calculations, the issue of
the center symmetry has
essentially  played no role in  analytical investigations so far.
In general, perturbative
calculations break this symmetry and thus inevitably are guided
towards the deconfined
phase. This is due to the change of the underlying gauge symmetry
from SU($N$) to
U(1)$^{N^{2}-1}$ when the coupling vanishes. In this limit,
the original
Z$_N$ symmetry gets effectively broken. For the center symmetry to
be preserved in the
path integral formulation, the Faddeev--Popov determinant \cite{Faddeev}
arising
in the process of gauge
fixing  cannot be treated perturbatively. In particular the
associated restrictions in the
range of integrations which prevent summation over Gribov
copies  cannot be neglected.
Likewise,
for the center symmetry to be preserved in the  canonical
formalism, the Gauss law has to be
resolved non-perturbatively. Only then the center symmetry
is guaranteed to appear as
the correct residual gauge symmetry.

For the formulation of the center symmetry and definition of
the Polyakov loops
we consider  QCD at finite extension, i.e., in a geometry where
the system
is of finite extent ($L$) in one direction
($x_{3}$), but of infinite extent in the other directions
(coordinates denoted by $x_{\perp} = (x_{0}, x_{1}, x_{2})$). Choosing
one compact coordinate
is of interest for additional  reasons. First, the parameter
$L$ helps to control infrared
ambiguities. This is of particular importance when using axial
like
gauges (for an early discussion of ambiguities in the axial gauge,
see
Ref. \cite{Schwinger}).
Second, by covariance, QCD at finite extension is equivalent to
finite
temperature QCD and, therefore, the essential properties of finite
extension QCD
are known from finite temperature lattice gauge calculations (for a recent
review, cf. \cite{lattrev}).
In particular the
presence of a phase transition at finite temperature implies,
via covariance,
occurrence of a phase transition when compressing the system,
i.e., decreasing
$L$. It also implies dimensional reduction to $2+1$ dimensional
QCD to occur \cite{Apple75,Apple81} if
the system is compressed far beyond the typical length scale
of strong
interaction physics. Although identification of the compact
direction with imaginary time
is more familiar, we will use the finite extension interpretation
by making a
spatial direction compact. In this case, the canonical
formulation is straightforward,
and the center symmetry plays the role of an ordinary
symmetry.

The variables of central importance in our investigation of the
center symmetry  are the Polyakov loops winding around the compact
3-direction.
These variables
characterize the phases of QCD (cf. \cite{Svetitsky} for finite
temperature QCD),
in particular the realization
of the center symmetry in the absence of quarks. Thereby they
serve as order
parameters of the confinement-deconfinement transition occuring
at a certain critical extension.

Besides the choice of the geometry, the choice of gauge is the
second important technical
ingredient on which our investigations are based. With one of
the space-time directions
made compact, the use of an axial type gauge seems natural.
Moreover in such a gauge
the Polyakov loop variables appear as elementary rather than
composite degrees of freedom. This makes the whole setting
particularly appropriate for a study of the
center symmetry and of the dynamics of the Polyakov loops.
Indeed the center
symmetry will  remain present at each level of the theoretical
development. This is possible only because
the elimination of redundant variables can be performed in closed form
without invoking  at any step perturbation theory.

Realization of the center symmetry implies the presence
of novel structural elements as compared to standard perturbative
QCD. In particular,  the center symmetric
phase with its
infinite energies associated with single static color charges does
not exhibit, as might be expected, the phenomenon of (chromo-electric) Debye
screening \cite{Debye}. Rather, naive application of perturbation theory
will be seen to
lead to
tachyonic behaviour of the Polyakov loop variables. This perturbative
instability indicates qualitative changes in the structure of
the vacuum which
will be shown to arise from the ultralocality of the Polyakov
loops. Ultralocality, i.e., the missing strength to generate wave phenomena,
represents an extreme form of confinement of these degrees of freedom. This
property will provide an appropriate framework for the discussion
of a variety
of non-perturbative, confinement related phenomena. On the other hand,
these drastic deviations in the
Polyakov loop vacuum from the perturbative one necessarily obscure the
standard short distance perturbative properties of QCD. This
dichotomy between
perturbative and non-perturbative physics will be illustrated
by a detailed discussion of the interaction energy
of static quarks. In this  approach to the Polyakov loop
dynamics, certain non-perturbative infrared properties are almost
trivial consequences of the
vacuum structure, while description of the short distance,
Coulomb-like
behaviour will be seen to require
coupling of
Polyakov loops to other gluonic variables to infinite order.

Aside from the Introduction and Conclusions, this paper is organized
into three main sections.
In Sect.~2, the QCD generating functional is derived
in a modified axial gauge, using a space with finite extension in the
3-direction.
Path integral quantization is used to rederive in a self-contained
and streamlined fashion results of our earlier canonical studies
(Sect.~2.2),
the main emphasis being on the SU(2) case (Sect.~2.3).
We have also included a brief
reminder on the equivalence of finite temperature and finite extension
field theory (Sect.~2.1) and illustrated the difficulties
encountered in the evaluation of screening effects, if one tries to
treat the Faddeev--Popov determinant perturbatively (Sect.~2.4).
Section 3 addresses the issue of Polyakov loops and exhibits ultralocality
as their main characteristics. In Sect.~3.2, we show how one can make
use of this property of the Polyakov loops to integrate them out,
thereby deriving an
effective theory for the other degrees of freedom which admits a
continuum limit. A qualitative discussion of the
phases of QCD (Sect.~3.3) and the confinement-deconfinement transition
(Sect.~3.4), as they appear in this novel description, follow.
Sect.~4 summarizes
our efforts to use the effective theory for investigating
properties of the center symmetric phase with the help of perturbation
theory, which now is markedly different from standard perturbation
theory. Feynman rules are given (Sect.~4.1) and the gluon two-point
function (Sect.~4.3) and Polyakov loop correlator at one (Sect.~4.2)
and two loop level (Sect.~4.4) are discussed in detail, emphasizing
the short distance aspects. These investigations are then used to
gain insight into the interaction between static quarks (Sect.~4.5) as
well as
shielding effects in the presence of dynamical quarks (Sect.~4.6).
The two appendices contain material of technical nature referred to in
the main text, namely the calculation of the electron self energy in
axial gauge QED (Appendix A) and the full expressions for the one loop gluon
self energy in modified axial gauge QCD (Appendix B).

\section{QCD in the Axial Gauge}

\subsection{Finite Extension versus Finite Temperature}
Before developing the formalism for axial gauge QCD we briefly
discuss the relation between QCD at finite extension and finite
temperature.
The equivalence of relativistic field theories at finite extension
and finite
temperature has been noted in Ref. \cite{Toms}
and used e.g. in a discussion of the finite
temperature quark propagators \cite{Koch}.
By rotational invariance in the Euclidean, the value of the
partition function of a system with finite extension $L$
in 3 direction and $\beta$  in 0 direction is invariant under the
exchange
of  these two extensions,
\begin{equation}
Z\left(\beta,L\right)=Z\left(L,\beta\right) \ ,
\label{FE1}
\end{equation}
provided bosonic (fermionic) fields satisfy periodic (antiperiodic)
boundary conditions in both time
and 3 coordinate.  Thus relativistic covariance connects the
thermodynamic properties of a canonical ensemble with the
properties of the pure state of the vacuum corresponding
to the same physical system but at finite extension. In particular,
as a consequence of (1), energy density and pressure are related
by
\begin{equation}
\epsilon\left(\beta,L\right)=-p\left(L,\beta\right) \ .
\label{FE2}
\end{equation}
For a system of non-interacting particles this relation connects energy
density or pressure of the Stefan Boltzmann law with the corresponding
quantities measured in the Casimir effect.

In QCD, by covariance, the existence of a phase transition at
finite temperature
implies the occurrence of a phase transition when compressing
the QCD vacuum (i.e., decreasing $ L$). From this point of view,
the confinement-deconfinement
phase transition or the chiral phase transition,
when quarks are
present, appear as ``quantum phase transitions'' (cf. \cite{YOUN95},
\cite{SGCS97}). They are driven by changes in quantum rather
than thermal fluctuations which in turn are induced by changes of a
parameter of the
system ($L$). Covariance connects quantitatively compressed
and heated systems
with each other. In particular, we conclude from Eq.~(\ref{FE2})
that in the
phase transition induced by compressing the system the Casimir
pressure changes
discontinuously, while the change in the energy density is continuous.
Furthermore, from results of lattice gauge calculations \cite{lattrev},
we infer
that this transition occurs
at a critical extension $L_{\rm c} \approx 0.8$ fm in the absence of
quarks and at $L_{\rm c} \approx 1.3$ fm when quarks are included.
These typical length scales indicate that we do not have to treat the
extension
strictly as an infrared parameter which tends to infinity. Rather we
expect no
essential property of the QCD vacuum to be changed significantly
if $L$ is of the
order of 2--3 fm. For extensions smaller than $L_{\rm c}$, the energy density
and pressure reach values which are typically 80 \% of the
corresponding  ``Casimir" energy and pressure. If the system is compressed
further and the extension becomes much smaller than  typical
length scales of
strong
interaction physics (e.g. a typical hadron radius $R$),
we expect correlation functions at transverse
momenta or energies $|p| \ll 1/L $ to be  dominated by the
zero ``Matsubara wave numbers" in 3-direction and thus to be given
by the dimensionally reduced QCD$_{2+1}$. Lattice calculations
\cite{Reisz} provide evidence for  this dimensional reduction
to occur if one of the Euclidean extensions becomes small.

The variables of central importance in our  investigation of
the role of the center symmetry  are the Polyakov loops winding
around the compact 3-direction
\begin{equation}
{\cal P} \left(x_{\perp}\right) = N_c^{-1} \mbox{tr}\,
\mbox{P}\exp\left\{ig \int_{0}^{L} dx_{3}  A_{3} \left(x\right)  \right\} .
\label{FE3}
\end{equation}
The vacuum expectation values of ${\cal P}$
characterizes the phases of QCD (cf. \cite{Svetitsky} for finite
temperature QCD), in particular their realization
of the center symmetry in the absence of quarks. Thereby they
serve as order parameters of the confinement-deconfinement
transition occurring when varying the extension.

\subsection{The Generating Functional}

For the theoretical treatment of QCD at finite extension or
finite temperature
an axial type gauge is particularly appropriate. In such gauges, the
associated Polyakov loops appear as fundamental rather than composite
degrees of freedom; this in turn permits a direct study of the dynamics of
the order parameter for the confinement-deconfinement transition. The
derivation of the axial gauge representation and a discussion of the
subtleties associated with this gauge choice has been carried out already
within the canonical formalism \cite{Lenz94,Lenz95}. Here we rewrite these
results into the path-integral formulation which serves as the basis
for the following developments (a related discussion is given in
Ref. \cite{Reinh}).
The formal expression for the QCD generating functional after gauge fixing
is
\begin{equation}
Z= \int d\left[A,\psi,\bar{\psi}\right] e^{i S_{\rm QCD}
\left[A,\psi,\bar{\psi}\right]}
\Delta_{\rm FP} \left[A\right]
\delta\left[f\left[A\right]\right]
\label{I1}
\end{equation}
with the standard QCD action
\begin{equation}
S_{\rm QCD} \left[A,\psi,\bar{\psi}\right] = \int d^{4}x \left\{-\frac{1}{4}
F^{a\mu\nu}F^{a}_{\mu\nu}
+\bar{\psi}\left(i/ \!\!\!\! D-m\right)\psi \right\} \ .
\label{I2}
\end{equation}
We require gluon and quark fields to satisfy periodic and antiperiodic
boundary conditions respectively,
\begin{eqnarray}
\label{I2a}
A_{\mu}^{a}\left(x_{\perp},x_{3}=L\right) & = &
A_{\mu}^{a}\left(x_{\perp},x_{3}=0\right) \ ,\nonumber\\
\psi\left(x_{\perp},x_{3}=L\right) & = &
\psi\left(x_{\perp},x_{3}=0\right) \ ,
\end{eqnarray}
with the notation
\begin{displaymath}
x=(x_{\perp},x_{3}) \ .
\end{displaymath}
This choice of boundary conditions is necessary for  the
equivalence of finite
extension and  finite temperature formulations.
Covariant derivative $D$ and field strength tensor $F$ are defined as usual,
\begin{eqnarray}
D_{\mu} & = & \partial_{\mu}+ig A_{\mu}                 \ ,
\nonumber\\
F_{\mu \nu } & = &
\partial_{\mu}A_{\nu}-\partial_{\nu}A_{\mu}+ ig\left[A_{\mu},
A_{\nu}\right]\ .
\label{I3}
\end{eqnarray}
Gauge fixing is implicitly  described by the  $\delta$-functional and the
corresponding Faddeev--Popov determinant in Eq.~(\ref{I1}) and is carried out
explicitly in two steps. First, the functional integral is constrained by
the following choice of $f$,
\begin{equation}
f_{x}^{a} \left[A\right]=A_{3}^{a} \left(x\right)\left(
1-\epsilon^{a}\right)+
\eta \partial_{3}A_{3}^{a}\epsilon^{a} \ ,
\label{I6}
\end{equation}
where $\epsilon^{a}$=1 (0) if $a$ refers to a diagonal (non-diagonal)
$\lambda$ matrix.
The Faddeev--Popov determinant associated with this gauge choice is determined
by
\begin{equation}
\delta\left[f\left[A\right]\right] {\cal M} \left(x,y;a,b\right)  =
-\delta\left[f\left[A\right]\right] \left( \partial _{3}^{y}\delta^{ab}
\left(1-\epsilon^{a}+\eta \epsilon^{a}\partial^{x}_{3}\right)-gf^{abc}
\epsilon^{c}A_{3}^{c} \left(y_{\perp}\right)\right) \delta^{4}
\left(x-y\right) \ .
\label{I7}
\end{equation}
The matrix ${\cal M}$ factorizes into contributions from ``neutral'' gluons
($\epsilon^{a}=\epsilon^{b}=1$) which are independent of the gauge fields
and from ``charged" gluons ($\epsilon^{a}=\epsilon^{b}=0$) which
give rise to a non-trivial Faddeev--Popov determinant det(${\cal M}$),
\begin{equation}
\delta\left[f\left[A\right]\right] {\cal M} \left(x,y;a,b\right)  =
- \delta\left[f\left[A\right]\right] \left( \partial _{3}^{y}\delta^{ab}-
g\epsilon^{c} f^{abc} A_{3}^{c} \left(y_{\perp}\right)\right)
\delta^{4} \left(x-y\right) \ .
\label{I9}
\end{equation}
The  Faddeev--Popov determinant is given by the product of eigenvalues
of the covariant derivative $D_3$
\begin{equation}
\left( \partial _{3}\delta^{ab}-g\epsilon^{c} f^{abc} A_{3}^{c}
\left(x_{\perp}\right)\right) \Psi^{b}\left(x \right) = i \mu \Psi^{a}
\left(x \right)
\label{I10}
\end{equation}
which can be calculated explicitly,
\begin{equation}
\mu_{p,q,n}\left(x_{\perp}\right)= \frac{2\pi n}{L}+
g\left( A_{3}^{qq} \left(x_{\perp}\right)-
A_{3}^{pp} \left(x_{\perp}\right)\right) \ ,
\label{I11}
\end{equation}
yielding
\begin{equation}
\Delta_{\rm FP}  =   \prod_{p,q,n,x_{\perp}}\mu_{p,q,n}\left(x_{\perp}\right)
\sim  \prod_{p>q,x_{\perp}} \sin^{2}\left[\frac{gL}{2} \left( A_{3}^{qq}
\left(x_{\perp}\right)- A_{3}^{pp} \left(x_{\perp}\right)\right)
\right] \ .
\label{I12}
\end{equation}
The gauge fixing by Eq.~(\ref{I6}) is not complete; Abelian, $x_{3}$
independent  gauge transformations leave the integrand in
Eq.~(\ref{I1}) invariant. In a second step a residual  gauge condition can be
imposed. For  perturbative applications a 2+1 dimensional Lorentz gauge
condition is implemented via a gauge fixing term in the action,
\begin{equation}
\label{I13}
S_{\rm gf}\left[A \right] =-\sum_{c_{0}=1}^{N-1}\int d^4x
\frac{1}{2 \xi} \left( \frac{1}{L}
\int_0^{L} dx_3 \partial^{\mu}A_{\mu}^{c_0} \right)^2 .
\end{equation}
The sum extends over the neutral gluons only. As in electrodynamics, no field
dependent  Faddeev--Popov determinant arises from this residual gauge fixing
(within the canonical formalism implementation of a residual Coulomb gauge
constraint is more natural, cf. \cite{Lenz94}).

Thus the generating functional of QCD in the axial gauge  can be written in
the following form,
\begin{equation}
Z=\int d\left[\psi,\bar{\psi}\right] \prod_{\mu=0}^{2} d\left[A_{\mu}\right]
\prod_{c_{0}=1}^{N-1}
d\left[A_{3}^{c_{0}}\right]\Delta_{\rm FP} \left[A\right]
e^{i(S \left[A,\psi,\bar{\psi}\right]+S_{\rm gf}\left[A \right])} \ .
\label{I16}
\end{equation}
The 3-components of the gauge fields have been eliminated up to 2+1
dimensional, neutral fields. Apart from the longitudinal, 2+1 dimensional,
neutral gluon fields  which appear in the gauge fixing term, no redundant
degrees of freedom are present anymore.

\subsection {QCD with SU(2) Color}
We analyse the formal structure of the generating functional
in the context of
SU(2)-QCD. This discussion will provide the basis for our dynamical
studies. In comparison to a naive
axial gauge formulation, the distinctive element in the generating functional
(\ref{I16}) is the presence of the Faddeev--Popov determinant (Eq.~(\ref{I12}))
which for SU(2) is given by
\begin{equation}
\Delta_{\rm FP}  =   \prod_{x_{\perp}} \sin^{2}
\left(gL A^{11}_{3}\left(x_{\perp}\right)\right) \ .
\label{I17}
\end{equation}
Gauge fields with polarization in the 3-direction cannot
be eliminated
completely; the eigenvalues of the Polyakov loops winding, for fixed
$x_{\perp}$, around the compact 3 direction are gauge invariant objects and
therefore have to be kept. The Faddeev--Popov determinant (Eqs.~(\ref{I12}),
(\ref{I17}))  is given by the Haar measure associated with these
particular group
elements which, in  SU(2), is the volume element of the first polar angle
in the parametrization of the group manifold by polar coordinates. It is thus
clear that the Faddeev--Popov determinant implicitly contains a
restriction to a
finite range of integration --- the fundamental domain determined
by the zeroes
of $\Delta_{\rm FP}$. For SU(2), this is the finite interval $[0,\pi]$
of definition  of the  polar angle $gLA_{3}^{11}$.

The presence of the Faddeev--Popov determinant with its restriction
in the range
of integration poses serious problems in defining the weak coupling limit
as the basis for perturbation theory. In the standard treatment,
the variables
$a_{3}$ are taken as Gaussian variables with the real axis as range
of definition, and the Faddeev--Popov determinant is effectively
neglected. Expansion of action and Faddeev--Popov determinant around $a_{3}=0$
is however problematic; at this point $\Delta_{\rm FP}$ vanishes and therefore
yields a singular contribution to the (effective) action. The meaning of
results obtained within such a framework such as  the electric
screening mass is not quite obvious. The subtleties of the weak coupling
limit in this gauge are related to the change in symmetry from SU($N$) to
U(1)$^{N^{2}-1}$ occurring at $g=0$. Concomitant
with this change in symmetry  is
a change from the $N-1$ gauge invariant eigenvalues of the Polyakov loops at
finite $g$ to the $N^{2}-1$ (U(1) gauge
invariant) ``zero-mode'' photons at $g=0$.
For the following it is convenient to introduce the
``Polyakov loop variables''
\begin{equation}
a_{3}\left(x_{\perp}\right)=  2 A_{3}^{11}\left(x_{\perp}\right)-
\frac{\pi}{gL}
\label{I18}
\end{equation}
and to redefine accordingly  charged gluon and quark fields,
\begin{eqnarray}
\label{I18a}
A_{\mu}^{pq}\left(x\right) & \rightarrow  & \exp{\left[-i x_{3}
\frac{\pi}{2L}
\left((\tau_{3})_{qq}-(\tau_{3})_{pp}\right)\right]} A_{\mu}^{pq}
\left(x\right) \ , \nonumber \\
\psi\left(x\right) & \rightarrow  & \exp{\left[-i x_{3}\frac{\pi}{2L}
\tau_{3}\right]}\psi\left(x\right) \ .
\end{eqnarray}
In this way, the  standard form of the action  is preserved and the
SU(2)-QCD generating functional given by
\begin{equation}
Z=\int  d\left[\psi,\bar{\psi}\right]\prod_{\mu=0}^{2} d\left[A_{\mu}\right]
d\left[a_{3}\right]\Delta_{\rm FP} \left[a_{3}\right]
e^{i(S \left[A_{\perp},a_{3},\psi,\bar{\psi}\right]+S_{\rm gf}\left[A_{\perp}
^{3}\right])}       \ ,
\label{I19}
\end{equation}
with the Faddeev--Popov determinant
\begin{equation}
\Delta_{\rm FP}\left[a_{3}\right]  =   \prod_{x_{\perp}} \cos^{2}
\left( gL a_{3} \left(x_{\perp}\right)/2\right)  \ .
\label{I20}
\end{equation}
For perturbative calculations  it might be advantageous to represent this
determinant as a functional
integral over  (anticommuting) ghost fields
\begin{equation}
e^{iS_{\rm FP}}=  \int d\left[ c,c^{\dagger}\right] e^{i S_{\rm gh}\left[c,
c^{\dagger}\right]}
\label{I21}
\end{equation}
with
\begin{equation}
S_{\rm gh}=\int d^{4}x c^{\dagger}_{a} \left(x\right) \left(\frac{1}{i}
\partial
_{3}\delta^{ab}+ig\epsilon^{3ab} a_{3} \left(x_{\perp}
\right)\right)c_{b} \left(x\right) \ .
\label{I22}
\end{equation}
Finally we observe  that due to the explicit $x_{3}$-dependence  of the
above field redefinitions and a similar treatment of the ghost fields,
changes in  the boundary
conditions occur,
\begin{eqnarray}
A_{\mu}^{a }\left(x_{\perp},x_{3}=L\right) & = & \left(-1\right)
^{1+\epsilon_{a}}A_{\mu}^{a}\left(x_{\perp},x_{3}=0\right) \nonumber\\
c_{a}\left(x_{\perp},x_{3}=L\right) &  =
& -c_{a}\left(x_{\perp},x_{3}=0\right)\nonumber \\
\psi\left(x_{\perp},x_{3}=L\right) & = &  e^{-i\pi\tau_{3} /2}
\psi\left(x_{\perp},x_{3}=0\right) \ ,
\label{I23}
\end{eqnarray}
i.e., neutral gluons remain periodic while charged gluons and ghosts satisfy
antiperiodic boundary conditions. Quark fields acquire a phase  $\pi/2$ in
going around the compact 3-direction. These changes in the boundary condition
will have important consequences. We emphasize at this point that
we have not modified the basic requirements of  periodicity or
antiperiodicity (Eq.~(\ref{I2a})) for boson and fermion fields, respectively.

In the axial gauge, with the Polyakov loops chosen to point in color 3
direction, charged and neutral gluons are  dynamically distinguished as
in any form of  ``Abelian  projection''. In such a formulation of QCD  it
often is convenient to use charged instead of cartesian color components,
\begin{equation}
\Phi_{\mu}\left(x\right)= \frac{1}{\sqrt{2}} \left(A_{\mu}^{1}\left(x\right)-
iA_{\mu}^{2}\left(x\right) \right)  \qquad  \mu=0,1,2  \ .
\label{I24}
\end{equation}
In terms of these charged gluon fields the  center (Z$_2$) symmetry
transformation which, in the absence of dynamical quarks,
leaves the generating
functional invariant
\begin{eqnarray}
a_{3}\left(x_{\perp}\right) & \rightarrow &-a_{3}\left(x_{\perp}\right)
\nonumber\\
C: \qquad \qquad A_{\mu}^{3}\left(x\right)
& \rightarrow & -A_{\mu}^{3}\left(x\right)
\label{J26} \\
\Phi_{\mu}\left(x\right) & \rightarrow & \Phi_{\mu}^{\dagger}\left(x\right)
\ ,  \nonumber
\end{eqnarray}
is the charge conjugation. We furthermore  note that the Polyakov loop is
given by
\begin{equation}
{\cal P} \left(x_{\perp}\right)= \mbox{tr}\, \mbox{P}
e^{ig \int_{0}^{L}dz A^{3}(x_
{\perp},z)}
= \sin\left(g L a_{3}\left(x_{\perp}
\right)/2  \right)
\label{I25}
\end{equation}
(in SU(2), ${\cal P}(x_{\bot})$ is hermitean so that there is no distinction
between
static quarks and antiquarks).
In the confined phase, at extensions $L>L_{c}$, charge conjugation
symmetry is realized
\begin{equation}
\label{I27}
C|0\rangle = \pm |0\rangle \ ,
\end{equation}
with  vanishing Polyakov loop expectation value. The ground state of
the deconfined phase at $L<L_{c}$ with its non-vanishing Polyakov loop
expectation value breaks spontaneously charge conjugation symmetry.

\subsection{Anti-Screening}
Before continuing with the development of the formalism we present a
preliminary perturbative analysis of screening properties associated
with Polyakov loops.
The relevant quantity to be calculated is the 33-component of the polarization
tensor $\Pi_{33}$ which, to lowest order and in the absence of dynamical
quarks, is given by the sum of tadpole, ghost and two gluon diagrams of
Fig.~1.
The intermediate charged  gluon propagators  of the tadpole (1a) and gluon
loop (1c)
diagrams and the ghost propagator of diagram (1b) are given by
\begin{eqnarray}
D^{ab}_{\mu\nu}\left(p\right) & = & \delta ^{ab}\frac{1}{p^2-p_{3}^2+i
\epsilon}\left[-g_{\mu\nu}+\frac{p_{\mu}p_{\nu}}{p_{3}^{2}}\right]\nonumber\\
\Delta^{ab}\left(p\right) &  = &  \delta ^{ab}\frac{1}{p_{3}} \ .
\label{I34}
\end{eqnarray}
The discrete values of the 3-component of the momenta are
\begin{equation}
p_{3,n}  =  \left(2n+1\right)\frac{\pi}{L} \ .
\label{I38}
\end{equation}
According to the ghost contribution to the action (Eq.~(\ref{I22})),
the ghost-Polyakov loop  vertex is given by
\begin{equation}
\label{I39}
V_{\rm gh P} = -g\epsilon^{3ab}
\end{equation}
with the color labels $a,b$ of the ghost fields.
Here and in the following, Greek indices denote the components 0,1,2, and
correspondingly we use the notation
\begin{displaymath}
p^{2}=p^{\mu}p_{\mu}= p_{0}^{2}- p_{1}^{2}-p_{2}^{2} \ .
\end{displaymath}
For our qualitative discussion of screening, we consider the simple case of
vanishing Polyakov loop momentum and obtain
\begin{equation}
\label{I40}
\Pi_{33}(0)=  2g^2 \frac{1}{L}\sum _{n=-\infty}^{\infty}  \int \frac{d^{3}p}
{\left(2\pi\right)^3}\left[m_{\rm tp}(p)+m_{\rm gh}(p)+m_{\rm gg}(p)\right]
\ ,
\end{equation}
with the following contributions from the 3 diagrams,
\begin{eqnarray}
\label{I41}
m_{\rm tp}(p)&=& -\frac{1}{ p_{3,n}^2}+ \frac{2}{p^2-p_{3,n}^2+i\epsilon}
\nonumber\\m_{\rm gh}(p)&=&  \frac{1}{ p_{3,n}^2}\nonumber\\
 m_{\rm gg}(p)&=& \frac{2}{ p_{3,n}^2}+ \frac{4p_{3,n}^2}
{\left(p^2-p_{3,n}^2+
i\epsilon\right)^2}
\end{eqnarray}
The integral in Eq.~(\ref{I41}) is performed with the help of dimensional
regularization,
\begin{equation}
\Pi_{33}(0)= \frac{i 2g^2}{\pi L}\sum _{n=-\infty}^{\infty} \sqrt{p_{3,n}^2}
\ ,
\label{I46}
\end{equation}
and the divergent sum is computed in $\zeta$-function regularization with the
final result
\begin{equation}
\Pi_{33}(0)= \frac{1}{3}\frac{i g^2}{ L^2} \ .
\label{I47}
\end{equation}
To appreciate the relevance of this result, we remark that in our
regularization procedure the ghost loop does not contribute at all;
the ghost propagator depends on the 3-component of the momentum
only and, in 3 dimensions,
the rules of dimensional regularization imply
\begin{equation}
\int \frac{d^{3}p}{\left(2\pi\right)^3} = 0    \ .
\label{I45}
\end{equation}
This result can be generalized to show that the ghost-gluon coupling has
no effect whatsoever to any order. We consider as an example the
diagram of Fig.~2.
Neither the ghost-Polyakov loop  vertices nor the propagators depend on the
(0,1,2)-components of the loop momentum $p$,  and therefore such
diagrams involving
ghost loops vanish. This suggests that no effects of the Faddeev--Popov
determinant are seen in perturbation theory, and one  might be tempted to drop
the corresponding contribution to the action in Eq.~(\ref{I19}).
In this case, one
could equally well try to define perturbation theory using the original
variable $A^{11}_3(x_{\perp})$. In our screening calculation the only change
concerns the values of the 3-components of the charged gluon momenta which
are now determined by periodic boundary conditions. Unfortunately a proper
definition of the charged gluon propagator is not possible due to the presence
of a zero mode in the  quadratic part of the charged gluon contribution to
the action (the gauge term in the first line of of Eq.~(\ref{I34})
is not defined for $p_{3}=0$).
Disregarding for the moment this difficulty and excluding 0 from the
momentum sum, we obtain the same formal expression for the polarization
tensor  as above with the momenta $p_{3,n}$ summed over the values
$2n\pi/L$. The final result is
\begin{equation}
\tilde{\Pi}_{33}(0)= -\frac{2}{3}\frac{i g^2}{ L^2}\ .
\label{I48}
\end{equation}
This result coincides, after Wick rotation,  with the standard value for the
squared Debye screening mass (cf. \cite{Debye,Weiss})
\begin{equation}
m_{\rm D}^{2} = \frac{2}{3} g^2T^2 ,  \label{I49}
\end{equation}
while the result of Eq.~(\ref{I47}) implies an imaginary value
for the screening mass.

Appearance of anti-screening apparently invalidates the perturbative approach
which led to this result. On the other hand,
the procedure that reproduces screening with the standard
value of $m_{\rm D}$ does not provide a viable alternative either. It is based
on a singular charged gluon propagator.
Source of these singularities is the change in symmetry which in turn
enforces a change in the number of zero modes of the associated differential
operators. While only one zero mode
corresponding to the covariant derivative $D_{3}$ exists at fixed
$x_{\perp}$, the spectrum of the
ordinary derivative $\partial_{3}$ of the U(1)$^{3}$ theory at $g=0$
contains three zero-modes. The difficulties encountered when treating the
Polyakov loop variables as Gaussian variables are not specific to our
particular approach. As is well known, when employing in finite temperature
perturbation theory the temporal gauge
$A_{0}=0$, ``spurious'' double poles appear which have to be
eliminated by additional prescriptions (cf. \cite{KAKA85}, \cite{JALA90},
\cite{LEST94}). Here, the origin of the difficulties,
is actually the elimination of physical variables, the Polyakov loop
variables  or, in the $g=0$
limit, the $N^{2}-1$ transverse photons. Even after implementation of
prescriptions for handling singularities, the resulting formalism remains
defective. Most importantly in the course of these manipulations, a QED like
shift symmetry (Z) associated with the Polyakov loop variables $a_{3}$
has been
introduced which is not present in the original theory. Thus one cannot
resolve the problems by resorting
to ambiguous or incomplete
gauge fixing
procedures. Rather the presence of anti-screening must be interpreted as a
dynamical failure of perturbation theory indicating instability of the
perturbative vacuum. In the following section we
shall show that the vacuum of Polyakov loops is that of ultralocal rather
than of Gaussian variables. Perturbation theory built upon this modified
vacuum
will turn out to be free of the above infrared problems.

\section{Dynamics of Polyakov loops}

\subsection{Ultralocal Polyakov Loops}

In this section we shall explicitly account for the non-Gaussian nature
of the Polyakov loop variables $a_{3}(x_{\perp})$ and respect the finite limit
of integration associated with these variables. To this end we first consider
the Polyakov loop dynamics in the absence of coupling to the other degrees of
freedom. The corresponding generating functional is, in the Euclidean,
given by
\begin{eqnarray}
Z_{0} & = & \int d\left[a_{3}\right]\Delta_{\rm FP} \left[a_{3}\right]
\exp \left\{-1/2 \int d^{4}x(\partial_{\mu} a_{3}(x_{\perp}))^{2}\right\}
\label{I50} \\
& = & \int_{-\pi/2}^{\pi/2} \prod_{x_{\perp}} d\tilde{a}_{3}\left(x_{\perp}
\right) \cos^{2}\tilde{a}_{3}\left(x_{\perp}\right)\exp \left\{-\frac{2\ell}
{g^{2}L}
\sum_{y_{\perp},\delta_{\perp}}(\tilde{a}_{3}( y_{\perp}+\delta_{\perp})-
\tilde{a}_{3}( y_{\perp}))^{2}\right\} \ .
\nonumber
\end{eqnarray}
We have discretized transverse space time, introduced the lattice spacing
$\ell$, lattice unit vectors $\delta_{\perp}$ and  have rescaled the Polyakov
loop variables
\begin{displaymath}
\tilde{a}_{3}(x_{\perp}) = gLa_{3}(x_{\perp})/2 \ .
\end{displaymath}
In the continuum limit,
\begin{equation}
\frac{\ell}{g^{2}L} \sim \frac{\ell}{L} \frac{1}{\ln \frac{\ell}{L}}
\rightarrow 0 \ ,
\label{I51}
\end{equation}
and therefore the nearest neighbour interaction generated by the Abelian
field energy of the Polyakov loop variables is negligible. As a consequence,
in the absence of coupling to other degrees of freedom, Polyakov loops do not
propagate,
\begin{equation}
\langle \Omega|T\left( a_{3} \left(x_{\perp}\right) a_{3} \left(0\right)
\right)|\Omega\rangle  \sim \left(\frac{\ell}{g^{2}L}\right)
^{x_{\perp}/\ell} \rightarrow
\delta^{3}
\left(x_{\perp}\right) \ .
\label{I52}
\end{equation}
Although the above procedure is similar to the strong coupling limit in
lattice gauge  theory, here we have not invoked a strong coupling
approximation. In the lattice dynamics of single links, the factor $1/g^{2}$
appears in the action and, as a consequence,  continuum limit and strong
coupling limit describe
two different regimes of the lattice theory. In the Polyakov loop dynamics on
the other hand which is controlled by the factor $\frac{\ell}{g^{2}L}$, strong
coupling and continuum limit coincide.

Non-flat measures for the Polyakov
loop variables with corresponding limited
ranges of integration appear in
gauge fixed formulations of QCD irrespective of space-time dimension and are
also important for the structure of the lower dimensional gauge fixed
theories. For 1+1 dimensional QCD with adjoint fermions e.g. it has been shown
\cite{Shifman} that only by accounting properly for the non-flatness
of the measure the symmetries of the system are correctly described. However
the specific dynamical consequences
of the compactness of these variables depend in general on the dimension of
space-time. It is interesting that the property
of ultralocality of the Polyakov loop variables seems to be unique
for 3+1
dimensions. As dimensional arguments dictate, the
relevant factor controlling the size of the action (cf. Eq.~(\ref{I50}))
is $\frac{1}{g^{2}L}$  in 2+1 dimensions,  implying
non-trivial
dynamics of the Polyakov loop variables $a_{3}$, and becomes
$\frac{1}{g^{2}\ell L}$ in 1+1 dimensions which exhibits the characteristic
dependence on the time slice of quantum mechanical variables.
Furthermore we remark that the above
result is a consequence of the finite range of integration of the Polyakov
loop variables. If we extend the range of definition to the full real
axis, the functional
\begin{equation}
Z_{0}^{\rm QED} = \int_{-\infty}^{\infty} \prod_{x_{\perp}} d\tilde{a}_{3}
\left(x_{\perp}\right) \cos^{2}\left(\tilde{a}_{3}\left(x_{\perp}\right)
\right)
\exp \left\{-\frac{2\ell}{g^{2}L}\sum_{y_{\perp},\delta_{\perp}}
(\tilde{a}_{3}( y_{\perp}+\delta_{\perp})- \tilde{a}_{3}( y_{\perp}))^{2}
\right\}
\label{I53}
\end{equation}
generates the ordinary Green functions of photons propagating in the 1-2
plane with polarization in the 3 direction (as shown above, after extending
the range of $a_{3}$ the presence of the Faddeev--Popov determinant is
irrelevant).

The ultralocality of the Polyakov loop variables is the basis of the
further developments and allows us to disregard the Abelian contribution
of $a_{3}$ to the action. We accordingly rewrite the generating functional
of Eq.~(\ref{I19}) as
\begin{eqnarray}
Z & = & \int  d\left[\psi,\bar{\psi}\right]\prod_{\mu=0}^{2}
d\left[A_{\mu}\right]
\exp \left\{ i(S \left[A_{\perp},\psi,\bar{\psi}\right]+S_{gf}\left[A_{\perp}
^{3}\right])\right\}\nonumber\\
&\cdot & \int d\left[a_{3}\right]\Delta_{\rm FP} \left[a_{3}\right]
\exp\left\{ i \int d^{3}x_{\perp} \left[g a_{3} \left(x_{\perp}\right)u
\left(x_{\perp}\right)+ g^2 a_{3}^{2} \left(x_{\perp}\right) v
\left(x_{\perp}\right)  \right] \right\} \ .
\label{I53a}
\end{eqnarray}
The composite field  $u\left(x_{\bot}\right)$ is  generated by the 3 gluon
interaction and the interaction of the Polyakov loops with quarks,
\begin{equation}
u \left(x_{\perp}\right) = \int_{0}^{L} dx_{3} \left\{i \Phi^{\dagger}_{\mu}
\left(x\right)\stackrel{\leftrightarrow}{\partial}_{3}  \Phi^{\mu}
\left(x\right)   - \bar{\psi} \left(x\right)  \frac{\tau_{3}}{2}\gamma_{3}
\psi\left(x\right)  \right\}                \ ,
\label{I54}
\end{equation}
while the field  $v\left(x_{\bot}\right)$  is generated by the 4 gluon
interaction,
\begin{equation}
v \left(x_{\perp}\right) = \int_{0}^{L} dx_{3} \Phi^{\dagger}_{\mu}
\left(x\right) \Phi^{\mu} \left(x\right) \ .
\label{I56}
\end{equation}
Eq.~(\ref{I53a}) is the essential result of this section and can serve on
the one hand as starting point for development of a Ginzburg--Landau theory for
the order parameter $a_{3}$  of the confinement-deconfinement transition.
In this case one has to formally integrate out the other degrees of
freedom.  On the other hand, one may integrate out the Polyakov loop variables
$a_{3}$. Here,  we shall  choose this alternative option and investigate
further the consequences of the peculiar  property of  ultralocality of the
Polyakov loop  variables.

\subsection{Effective Action and Order Parameter}

\label{sec:2}

In this section we integrate out explicitly the Polyakov loop variables and
derive the generating functional for gluon ($A_{\mu},\ \mu=0,1,2 $) and quark
Green functions; in this way we also will be able to arrive at a novel
representation of the Polyakov loop expectation value and the associated
correlation function. To this end we expand the exponential in
Eq.~(\ref{I53}) and  keep only the leading term in a $l/(g^2L)$
expansion,
\begin{eqnarray}
\int d\left[a_{3}\right]&\Delta_{\rm FP} \left[a_{3}\right]&
\exp\left\{i \int d^{3}x_{\perp} \left[g a_{3} \left(x_{\perp}\right)
u \left(x_{\perp}\right) + g^2 a_{3}^{2} \left(x_{\perp}\right)
v \left(x_{\perp}\right)  \right]\right\}
\nonumber \\
& \approx &  \prod_{\vec{x}_{\perp}}\left[  1+i \frac{\ell^{3}}{L^{2}}
\left(\frac{\pi^2}{3}-2\right) v \left(\vec{x}_{\perp}\right)\right]\ .
\label{J1}
\end{eqnarray}
This expansion is justified if  the fluctuations of
the gluon fields are controlled by the ultraviolet cutoff $1/\ell$ as in a
non-interacting theory, and if a possible violation of the reflection symmetry
($x_{3} \rightarrow -x_{3}$) is limited to finite  momenta
($|p_{3,n}|<\Lambda \ll \frac{1}{\ell}$),
\begin{displaymath}
A_{\mu} \sim \frac{1}{\ell}\quad \mbox{i.e.}\quad v\sim \frac{L}{\ell^{2}}
\ , \quad u\sim \frac{\Lambda L}{\ell^{2}} \ .
\end{displaymath}
We also note that in this expansion, we disregard non-perturbative dynamics
associated
with singular field configurations. We thus rewrite the generating
functional as
\begin{equation}
Z =  \int d\left[A, \psi,\bar{\psi}\right]  e^{ i S_{\rm eff}
\left[A,\psi,\bar{\psi}\right] }
\label{J2}
\end{equation}
with the effective action given by
\begin{equation}
\label{J3}
S_{\rm eff} \left[A,\psi,\bar{\psi}\right]=
\int d^{4}x{\cal L}_{\rm eff}  =
S\left[A,\psi,\bar{\psi}\right]+
S_{\rm gf}\left[A_{\perp}^{3}\right]+\frac{1}{2} M^{2}\sum_{a=1,2}
\int d^{4}x A^{a}_{\mu} \left(x\right)  A^{a,\mu} \left(x\right) \ .
\end{equation}
Expectation values and correlation functions of the Polyakov loops are
calculated correspondingly with the following results,
\begin{eqnarray}
\langle \Omega|\sin(gL a_{3} \left(x_{\perp}\right) /2 )|\Omega\rangle
& \approx &
\frac{1}{Z\left[0\right]}
\frac{16 i\ell^{3}}{9 \pi L}   \int d\left[A, \psi,\bar{\psi}\right]
u\left(x_{\perp}\right)   e^{iS_{\rm eff} \left[A,\psi,\bar{\psi}\right]}
\nonumber\\
& \sim & \langle \Omega| u\left(x_{\perp}\right)   |\Omega\rangle \ ,
\label{J4}
\end{eqnarray}
\begin{equation}
\langle \Omega|T\left(\sin (gL a_{3} \left(x_{\perp}\right)/2) \sin (gL a_{3}
\left(y_{\perp}\right)/2)\right)|\Omega\rangle \quad  \sim  \quad \langle
\Omega|T
\left[u\left( x_{\perp}\right) u\left( y_{\perp}\right)\right]|\Omega
\rangle\  .
\nonumber \\
\label{J5}
\end{equation}
The effective action of QCD after integrating out the Polyakov loops
(Eq.~(\ref{J3})) is that
of QCD in the naive axial gauge ($A_{3}=0$) complemented by a lower
dimensional residual gauge fixing term, a mass term and the already discussed
change in boundary conditions of charged gluon and quark fields. In the spirit
of the Abelian projection we write the effective Lagrangian as
\begin{eqnarray}
{\cal L}_{\rm eff}&
=- &\frac{1}{2} \left(d_{\mu}^{*}\Phi_{\nu}^{\dagger}-d_{\nu}^{*}
\Phi_{\mu}^{\dagger}\right) \left(d^{\mu}\Phi^{\nu}-d^{\nu}\Phi^{\mu}\right)
-\partial_{3}\Phi_{\mu}^{\dagger}\partial^{3}\Phi^{\mu}
+M^{2}\Phi_{\mu}^{\dagger}\Phi^{\mu}\nonumber\\
&-& \frac{1}{4}f_{\mu \nu}f^{\mu \nu} -\frac{1}{2}\partial_{3}A_{\mu}
\partial^{3}A^{\mu} -\frac{1}{2\xi } \left(\frac{1}{L}\int_{0}^{L} dx_{3}
\partial ^{\mu}A_{\mu}\right)^{2} \nonumber\\
&+&  ig f^{\mu \nu} \Phi^{\dagger}_{\mu}\Phi_{\nu}-\frac{g^{2}}{4}
\left(\Phi^{\dagger}_{\mu}\Phi_{\nu}-\Phi^{\dagger}_{\nu}\Phi_{\mu}\right)
\left(\Phi^{\nu \dagger}\Phi^{\mu}-\Phi^{\mu \dagger}\Phi^{\nu}\right)
\nonumber\\
& + & \bar{\psi}\left(i / \!\!\! d -m\right)\psi  +  \bar{\psi}i
\gamma^{3}\partial_{3}\psi  -\frac{g}{\sqrt{2}} \bar{\psi}\gamma^{\mu}
\left(\Phi_{\mu}\tau_{+}+ \Phi_{\mu}^{\dagger}\tau_{-} \right)\psi \ .
\label{J6}
\end{eqnarray}
In addition to the charged gluon fields we have introduced Abelian field
strengths  generated by the neutral gluons,
\begin{equation}
f_{\mu \nu} = \partial_{\mu}A_{\nu}- \partial_{\nu}A_{\mu}  \ ,
\quad A_{\mu} = A_{\mu}^{3}\ ,
\label{J7}
\end{equation}
and the corresponding covariant derivatives
\begin{equation}
d_{\mu}= \partial_{\mu} +ig A_{\mu} \ , \qquad \  / \!\!\! d = \gamma^{\mu}
\left( \partial_{\mu} +ig A_{\mu}   \frac{\tau_{3}}{2}\right) \ .
\label{J8}
\end{equation}
The Polyakov loops leave the antiperiodic boundary conditions and the mass
term of the charged gluon fields as their signatures.
The antiperiodic boundary conditions reflect the mean value of the Polyakov
loop variables, the geometrical mass their fluctuations. Clearly the precise
value of this geometrical mass
\begin{equation}
M^{2}=\left(\frac{\pi^{2}}{3}-2\right) \frac{1}{L^{2}}
\label{J9}
\end{equation}
depends on the particular form of the Faddeev--Popov determinant. The emergence
of this geometrical  mass with its characteristic independence of the
coupling constant is a consequence of the finite range of integration.
Unlike  mass generation, the change in boundary conditions is a less
common phenomenon. For its interpretation we observe that the transformation
to antiperiodic charged gluon variables is a merely formal device.
The antiperiodic boundary conditions in
(\ref{I23}) actually describe the appearance of Aharonov--Bohm fluxes
in the elimination of the Polyakov loop variables. Periodic
charged gluon fields may be used if the differential operator
$\partial_{3}$
is replaced by
\begin{equation}
\partial_{3} \rightarrow \partial_{3}+\frac{i\pi}{2 L}[\tau_{3},
\qquad .
\label{J10}
\end{equation}
As for a quantum mechanical particle on a circle, such a
magnetic flux is technically most easily  accounted for by an
appropriate  change in boundary conditions --- without changing
the original periodicity requirements. With regard to the rather
unexpected physical consequences, the space-time independence of
this flux is important, since it induces global changes in the theory.
These global changes are missed if the Polyakov loops  are treated as
Gaussian variables.
Thus expressed in these more physical terms, the charged gluons are
massive and move in a constant
color neutral gauge field pointing in the spatial 3 direction of the strength
$ \frac{\pi}{g L}$. Since $x_{3}$ is a compact variable  we can
associate a color magnetic flux  with this gauge field,
\begin{equation}
\Phi_{\rm mag}= \frac{\pi}{g} \ .
\label{J12}
\end{equation}
The corresponding magnetic field of strength
\begin{equation}
B= \frac{1}{g L^{2}}
\label{J13}
\end{equation}
lives however in the unphysical embedding space ---  e.g. in the interior of the
cylinder whose surface is the spatial manifold of QCD$_{2+1}$.

Summarizing the results of this section, we emphasize the crucial property of
ultralocality of the Polyakov loops. It implies that these variables do not
constitute physical degrees of freedom; rather they are dependent variables.
In the axial gauge these dependent variables are composite gluon
fields. Propagation of the Polyakov loops occurs only via intermediate
excitation of ``two gluon states'' created by the operator $u(x_{\perp})$. In
the course of integrating out these dependent variables, dynamical differences
between neutral and charged gluons arise. Charged gluons
acquire a mass and are subject to antiperiodic boundary conditions while
neutral gluons remain unaffected at the perturbative level. It appears
that with these dynamical differences, the formalism contains the
seeds for Abelian dominance of long distance physics. Results of recent
lattice
calculations of the gluon
propagator in maximally Abelian gauge have actually been interpreted in terms
of massive charged and essentially massless neutral gluons \cite{AMSU97}.

\subsection{Phases of QCD in Axial Gauge}

The following discussion  will focus on the properties of the SU(2) Yang Mills
theory. If dynamical quarks are present, the charge
conjugation (Eq.~(\ref{J26}))
is not a symmetry transformation of the system. Formally, the Lagrangian of
Eq.~(\ref{J6}) remains invariant if the quark fields too are transformed,
\begin{equation}
\psi \rightarrow  \tau_{1}\psi  \ .
\label{SP18}
\end{equation}
This transformation changes however the boundary conditions (Eq.~(\ref{I23})).
In the absence of dynamical quarks, the auxiliary field $u(x_{\perp})$ which
is odd under charge conjugation
\begin{equation}
C:\quad \quad u \left(x_{\perp}\right) \rightarrow  -u \left(x_{\perp}\right)
\label{SP20}
\end{equation}
serves, after elimination of the Polyakov loop variables, as an order
parameter for the realization of charge conjugation symmetry.
In the confined phase, charge conjugation is realized
\begin{equation}
\langle 0| u \left(x_{\perp}\right) |0\rangle  = 0 \quad ,\quad L> L_c
\label{SP21}
\end{equation}
and spontaneously broken in the deconfined phase
\begin{equation}
\langle 0| u \left(x_{\perp}\right) |0\rangle  \neq  0 \quad ,\quad L< L_c \ .
\label{SP22}
\end{equation}
The deconfinement transition occurring when decreasing $L$ beyond the
critical value is accompanied or possibly generated by color currents in
the compact 3-direction. The presence of these currents signals a
simultaneous breakdown of the reflection symmetry $x_{3} \rightarrow  -x_{3}$
as implied  by the non-vanishing vacuum expectation value of $u$.

The perturbative ground state of QCD is symmetric under charge conjugation,
i.e., it respects the center symmetry. This is the distinctive property of QCD
in the (modified) axial gauge. Unlike ``perturbative'' gauge choices such as
covariant gauges or the Coulomb gauge in the standard treatment
which incorporate the symmetries of electrodynamics (the U(1)$^{N^{2}-1}$
theory), the modified  axial gauge with its  non-perturbative resolution of
Gauss' law preserves the Z$_N$ symmetry characteristic for Yang Mills
theory. By respecting the charge symmetry, the perturbative vacuum
satisfies the confinement criterium and indicates an infinite energy to be
associated with a static fundamental charge. Needless to say, the
perturbative limit is insufficient to describe realistically the phenomena
related to confinement; it however appears that certain  global properties of
the system are properly accounted for already at this (modified) perturbative
level, leading naturally into the confined phase of QCD.

The correlation function corresponding to this auxiliary field provides
further characterization of the phases of QCD. After rotation to
Euclidean time
\begin{equation}
x_{0} \rightarrow  -i x_{0}^{E}
\label{SP23}
\end{equation}
the correlation function of the Polyakov loops  yields the interaction energy
$V$ of static color charges (in the fundamental representation). Thus we have
after adjusting an additive
constant in $V$ which accounts for the proportionality factor in Eq.~(\ref{J5})
\begin{equation}
\exp{\left\{-LV\left(r \right)\right\}}= \langle \Omega|T \left[u\left(
x^{E}_{\perp}\right) u\left(0\right)\right]|\Omega\rangle  = {\cal D}(r)
\ ,
\qquad  r^{2}= \left(x^{E}\right)^{2} .
\label{SP24}
\end{equation}
Due to the rotational invariance in Euclidean transverse space, we are free to
choose  $x^{E}$ to point in the time direction. We insert a complete set of
excited states
\begin{equation}
\exp{\left\{-L V\left(r \right)\right\}}= \sum _{n}\left|
\langle n|u\left(0\right)|\Omega\rangle  \right|^{2}e^{-E_{n}  r} \ .
\label{SP25}
\end{equation}
In the confined phase, the ground state does not contribute to this sum
(cf. Eq.~(\ref{SP21})). If the spectrum exhibits a gap, the potential energy
$V$ increases linearly with $r$ for large separations,
\begin{equation}
V\left(r \right) \approx \frac{E_{1}}{L} r \quad  \mbox{for}\quad r
\rightarrow \infty \quad \mbox{and} \quad L > L_c \ .
\label{SP26}
\end{equation}
Since on the other hand, the slope is given by the string tension  $\sigma$,
we  conclude that the spectrum of states excited by the composite
operator $u$  possesses a gap which increases linearly with the extension $L$.
Thus in  Yang Mills theory at finite extension  the phenomenon
of confinement is connected to a shift
in the spectrum of gluonic excitations to excitation energies
\begin{equation}
E \geq  \sigma L \ .
\label{SP27}
\end{equation}
Note that the  class of states
excited by the Polyakov loop variables are associated in QED with a vanishing
threshold energy.
This result implies in particular that in the confined phase, glueball states
which, for sufficiently large values of $L$, should not be affected by the
finite extension, cannot be excited by  $u$. The characteristic property of
the ``two gluon'' operator $u$ which most likely is  responsible
for this confinement  phenomenon  is the negative $C$-parity.
It is remarkable that  perturbation theory yields the linear rise of the static
quark-antiquark potential at large separations. The charge symmetric ground
state does not contribute to the
sum in Eq.~(\ref{SP25}), and  the spectrum of charged gluons exhibits a
gap as a consequence of the antiperiodic boundary conditions and the
geometrical mass term. The resulting $L$-dependence of the  perturbative
string tension
\begin{equation}
\label{SP28}
\sigma_{\rm pert} = \frac{2}{L}\sqrt{M^{2}+(\pi/L)^{2}}
\end{equation}
is, at this point, determined by dimensional arguments and obviously not
realistic.

A further characterization of the negative and positive $C$-parity sectors of
the Z$_2$ symmetric phase can be obtained through a discussion of the adjoint
Polyakov loops. The adjoint Polyakov loop is defined with the matrices
$T^{a}$ of the adjoint representation as
\begin{equation}
{\cal P}_{\rm ad}(x_{\perp}) = \frac{1}{3} \mbox{Tr}\,{\mbox P}
\exp \left(ig\int_{0}^{L}
dz A_3^{a}(x_{\perp},z)  T^{a}\right)
\label{SP27a}
\end{equation}
and, if expressed in terms of the variables  $a_{3}$ of Eq.~(\ref{I18}),
given by
\begin{equation}
\label{SP27b}
{\cal P}_{\rm ad}(x_{\perp})=\frac{1}{3} \left(1-2\cos gLa_{3}(x_{\perp})
\right)\ .
\end{equation}
Functional  integration over the ultralocal Polyakov loop variables is performed
as in Sect.~3.2 (cf. Eqs. (\ref{J3}) -- (\ref{J5})) and yields expressions of
expectation value  and associated correlator of the adjoint Polyakov loop
in terms  of the composite field $v(x_{\perp})$ of Eq.~(\ref{I56})
\begin{equation}
\label{SP27c}
\langle \Omega |{\cal P}_{\rm ad}|\Omega \rangle  \sim
\langle \Omega|  v |\Omega \rangle \ , \qquad
\langle \Omega |T  \left[ {\cal P}_{\rm ad}(x_{\bot}) {\cal P}_{\rm ad}(0)
\right] |\Omega
\rangle
\sim \langle \Omega |T  \left[ v(x_{\bot}) v(0) \right] |\Omega \rangle \ .
\label{SP27d}
\end{equation}
Unlike the order parameter $u(x_{\perp})$, the field  $v(x_{\perp})$ has
positive $C$-parity and is therefore not prevented by a selection rule from
acquiring an  expectation value. Indeed already at the perturbative level
such an
expectation value occurs, given by the tadpole contribution to the
$a_{3}$ effective action generated by the corresponding 4-gluon vertex. As a
consequence of  the non-vanishing expectation value, the
interaction energy between static adjoint charges
\begin{equation}
\exp{\left\{-L V_{\rm ad}\left(r \right)\right\}}= \sum _{n}\left| \langle n|v
\left(0\right)|\Omega\rangle  \right|^{2}e^{-E_{n}  r}
\label{SP27e}
\end{equation}
decreases exponentially  at large  distances,
\begin{equation}
V_{\rm ad}\left(r \right) \sim \frac{1}{r^2} e^{-E_{1} r}  \ ,
\label{SP27f}
\end{equation}
with $E_{1}$ the threshold of excited states. On the other hand,  we expect
quite generally this exponential decrease to be determined by the lowest
glueball mass. Thus, unlike the two gluon operator $u(x_{\perp})$, the composite
operator $v(x_{\perp})$ yields excitations in the physical sector of hadronic
states.

\subsection{Confinement-Deconfinement Transition}
The perturbative Z$_2$ symmetric phase of QCD reached in this modified axial
gauge not only shares characteristic  properties with the
non-perturbative confining phase, it also exhibits signatures which point to the
necessity of a phase transition to the deconfined phase with the Z$_2$ symmetry
spontaneously broken. We start with a discussion of the ground state energy
density. The total energy of the system is given by
\begin{equation}
E= L_{\perp}^{2}\int \frac{ d^{2} k}{\left(2 \pi\right)^{2}}
\sum _{n=-\infty}^{\infty}\left[ \left( k^{2}+
\frac{\pi^{2}}{L^2} \left(2n\right)^{2}\right) ^{1/2}+ 2\left( k^{2}+
M^{2}+
\frac{\pi^{2}}{L^2} \left(2n+1\right)^{2}\right) ^{1/2} \right],
\label{CE0}
\end{equation}
with $L_{\perp}$ denoting the extension of the system in the transverse (1,2)
directions. The two terms represent the zero point energies of
neutral and charged gluons, respectively. For technical simplicity of this
qualitative discussion  we shall neglect the charged gluon mass and  compute
the ground state energy with the help of dimensional regularization. It is
straightforward to show
\begin{equation}
\mu^{2-2\omega} \int \frac{d^{2 \omega}k}{\left(2 \pi\right)^{2 \omega}}\sum_
{n=-\infty}^{\infty} \left( k^{2}+ \frac{1}{L^{2}} \left(2\pi n
+\chi\right)^{2}\right)^{1/2} = - \frac{\pi^{2}}{45 L^{3}}
\left[ 1-\frac{15}{8}
\left( \left(\frac{\chi}{\pi}-1\right)^{2}-1\right)^2\right]
\label{CE1}
\end{equation}
in the limit  $\omega \rightarrow 1$.
This result implies  the standard Casimir energy or Stefan Boltzmann law for
the neutral gluons ($\chi=0$)
\begin{equation}
\label{CE2}
E_{\rm neut}/( L_{\perp}^{2}L) = - \frac{\pi^{2}}{45 L^{4}}
\end{equation}
while (up to an important sign) the expression for the charged gluon
energy ($\chi=\pi$) is reminiscent
of the fermionic contribution to the energy density
\begin{equation}
\label{CE2a}
E_{\rm ch}/( L_{\perp}^{2}L) =  \frac{7}{4} \frac{\pi^{2}}{45 L^{4}} \ .
\end{equation}
The total ground state energy is thus given by
\begin{equation}
\label{CE3}
E_{\rm ch}/( L_{\perp}^{2}L) =  \frac{3}{4} \frac{\pi^{2}}{45 L^{4}}\ .
\end{equation}
Due to the antiperiodic boundary conditions of the charged gluons a
change of sign occurs in the ground state energy which, in turn, implies a
change of sign in the pressure, i.e., a repulsive Casimir force acting  between
the plates enclosing the system. Invoking covariance, this change in the
characteristic properties of the Casimir effect is seen to lead to a change
of sign in the relevant thermodynamic properties of the same system at
infinite extension but finite temperature. Thus the perturbative Z$_2$ phase
is  thermodynamically unstable. This instability is of little relevance for
the large extension or
low temperature phase, where the  non-perturbative phenomena of
confinement and  generation of a mass gap in the
``hadronic'' sector will change  the power law in the Casimir energy and
pressure into an exponential dependence. Likewise, the appearance of the
imaginary ``screening mass'' in the polarization
propagator discussed above (cf. Eq.~(\ref{I47})) which signals this instability
poses no problems   given the zero-range of the Polyakov loop propagator (see
below for a more detailed discussion).
At small extension
or high temperature on the other hand, where a perturbative approach
should be appropriate, this
instability seems to rule out a Z$_2$ symmetric high temperature
phase. Such a phase indeed would have properties very different from the high
temperature phase as deduced from lattice gauge calculations. Trivially,
with  the center symmetry realized, such a phase would  have to exhibit certain
characteristics of confinement. Furthermore, irrespective of the dynamics, at
high temperatures dimensional reduction should take place. Like quarks in QCD,
the charged  gluons decouple
from the low-lying excitations due to their antiperiodic boundary conditions
in the process of dimensional reduction. Thus at small  extension or high
temperature, the Z$_2$ symmetric  phase is described by QED$_{2+1}$ rather than
by QCD$_{2+1}$.

From our discussion the following qualitative,  axial gauge description of
the confinement-deconfinement transition emerges. After a gradual decrease in
the threshold of states with  negative $C$-parity  with decreasing $L$,
the whole spectrum of excitations ($C=\pm 1$) becomes suddenly available
when at the deconfinement transition with the breakdown of $C$-parity
simultaneously   string tension (threshold of the $C=-1$ states) and  mass gap
in the hadronic sector
(threshold of $C=1$ states) vanish. In this transition,
the charged gluon fields effectively must become periodic
(up to possible interaction effects), i.e., the Aharonov--Bohm fluxes
(Eq.~(\ref{J12})) are  shielded
and simultaneously  the geometrical mass  $M$ (Eq.~(\ref{J9})) disappears.
As a result of the  phase
transition, the unlimited  increase in the lowest  single gluon
energy
\begin{equation}
\epsilon (\vec{k}_{\perp},n_{3})^{2} = {k}_{\perp}^{2}+
\frac{1}{L^{2}} \left( \pi^{2}\left(2 n_{3}+1 \right)^{2} +
\left(\pi^{2}/3-2\right)\right).
\label{CE4}
\end{equation}
with  decreasing extension $L$  is  prevented and
thereby, in dimensional reduction, the correct high temperature limit
is reached.
In this change of boundary conditions, the degeneracy of oppositely
charged gluons
with momenta $n_{3}=0$ and $n_{3}=-1$  is lifted and currents
in the 3-direction
(cf. Eqs. (\ref{I54}), (\ref{SP22})) are generated.
Finally,  this change of  boundary conditions results in a change
in Casimir energy density and pressure which according to Eq.~(\ref{CE1})
is given by
\begin{equation}
\Delta \epsilon=  - \pi^{2} /12 L^4 \ ,\qquad \Delta p =
3\Delta \epsilon \ . \label{FE18}
\end{equation}
This estimate is of the order of magnitude of the change in the energy
density across the confinement-deconfinement transition when compressing
the system,
\begin{equation}
\Delta \epsilon=  -0.45 /L^4 \ ,
\label{FE19}
\end{equation}
deduced from the finite temperature lattice calculation of
Ref. \cite{Engels}.

In summary, the thermodynamic instability of the Z$_2$ symmetric perturbative
phase implies the presence of non-perturbative phenomena to stabilize this
phase and a transition to a phase with broken Z$_2$ symmetry.
If  QCD in the high temperature (or small extension) deconfined
phase is to be described perturbatively, one therefore has to abandon
this modified axial or  temporal  gauge with its characteristic $N-1$
Polyakov loops as zero modes. Starting point has to be  QCD at $g=0$, and
for this  U(1)$^{N^2-1}$ theory an axial gauge with $N^2-1$ photons as
zero modes is more appropriate. This procedure breaks the center symmetry
and the Stefan--Boltzmann law at  high temperatures is guaranteed
by gauge choice.

\section{Perturbation Theory in the Z$_2$ Symmetric Phase}
\label{sec:3}

\subsection{Feynman Rules}

In this section we shall continue our analysis of the  Z$_2$ symmetric phase
with a discussion of specific issues in perturbation theory. In this context
the perturbative treatment of the Polyakov loop correlator will be of particular
importance. The Feynman rules are easily derived from the quadratic part of
the effective Lagrangian of Eq.~(\ref{J6}). As a result of integrating out  the
Polyakov loop variables, charged and neutral gluon propagators have different
momentum dependences. The charged gluon propagator is given by
\begin{equation}
D_{\mu \nu}^{ab}(p,p_3) = \frac{\delta^{ab}(1-\delta^{a3})}
{p^2-p_3^2-M^2+i\epsilon}
\left[-g_{\mu \nu} + \frac{p_{\mu} p_{\nu}}{p_3^2+M^2} \right]
\label{q1}
\end{equation}
with $p_3$ derived from antiperiodic boundary conditions,
\begin{equation}
p_3= \frac{2\pi}{L} \left(n +\frac{1}{2}\right)  \  ,
\label{q2}
\end{equation}
and the neutral gluon propagator by
\begin{equation}
D_{\mu \nu}^{ab}(p,p_3) = \frac{\delta^{ab}\delta^{a3}}{p^2-p_3^2+i\epsilon}
\left[-g_{\mu \nu} + p_{\mu} p_{\nu}\left((1-\delta_{p_3,0})\frac{1}{p_3^2}
+\delta_{p_3,0}(1-\xi)\frac{1}{p^2+i\epsilon}\right) \right]
\label{q3}
\end{equation}
with $p_3$ derived from periodic boundary conditions,
\begin{equation}
p_3= \frac{2\pi n}{L} \ .
\label{q4}
\end{equation}
In the actual calculations, the parameter $\xi$ of the residual covariant
gauge will be set equal to 1 (2+1 dimensional Feynman gauge).
The 3- and 4-gluon vertices are standard,  except that only three
polarizations (0,1,2) appear, and given by
\begin{equation}
V^{abc}_{\lambda\mu\nu}(p,q,r) =  g \epsilon^{abc} \left[ (r-q)_{\lambda}
g_{\mu \nu} +(q-p)_{\nu}
g_{\lambda \mu} + (p-r)_{\mu} g_{\nu \lambda} \right]
\delta^{(3)}(p+q+r) \delta_{p_3+q_3+r_3,0}
\nonumber
\end{equation}
\begin{eqnarray}
W^{abcd}_{\lambda\mu\nu\rho}(p,q,r,s) &=& -ig^2 \left[ \epsilon^{fab}
\epsilon^{fbc} \left( g_{\lambda \nu}g_{\mu \rho}
-g_{\lambda \rho} g_{\mu \nu} \right) + \epsilon^{fcb}\epsilon^{fad}
\left(g_{\lambda \nu} g_{\mu \rho} - g_{\lambda \mu} g_{\nu \rho} \right)
\right.
\nonumber \\
& & \left. + \epsilon^{fac}\epsilon^{fbd} \left( g_{\lambda \mu} g_{\nu \rho}
- g_{\lambda
\rho} g_{\mu \nu} \right) \right] \delta^{(3)}
(p+q+r+s) \delta_{p_3+q_3+r_3+s_3,0}
\label{q4a}
\end{eqnarray}
with $a,\lambda ,p$ etc. denoting color, polarization and
momentum of incoming
gluons. Finally, the coupling of a Polyakov loop to two gluons has the form
\begin{equation}
V_{{\rm ggP}}=V^{3bc}_{3\mu\nu}(p_{\perp},q,r)= g \epsilon^{3bc} (r-q)_3
g_{\mu \nu} \delta^{(3)}(p+q+r) \delta_{q_3+r_3,0} \ .
\label{q4b}
\end{equation}
We observe that both neutral and charged gluon propagators are well defined in
the infrared. By properly accounting for the zero modes of the differential
operators appearing in the quadratic part of the action no infrared infinities
are encountered. The antiperiodic boundary conditions and the geometrical mass
term yield a well defined charged gluon propagator while as in QED, the
residual gauge fixing is instrumental for the proper infrared behavior of
the neutral gluon propagator. Thus the characteristic difficulties of
the continuum axial gauge propagator, such as the appearance of spurious
double poles \cite{Landshoff}, are not present.
The infrared properties are also
significantly different from those obtained in standard finite temperature
formulation. In our case, infrared
infinities can occur only in loops if they are generated by  4-gluon vertices
with vanishing external momenta and if both gluons in the loop are
neutral. Thus the infrared properties resemble those of scalar QED and
the difficulties encountered in finite temperature perturbation
theory for QCD (cf. \cite{LIND80}) can
therefore be expected to be alleviated in
the Z$_2$ symmetric phase.

The following discussion will focus on the properties of the Polyakov loop
correlation
function which in Eq.~(\ref{J5}) has been expressed in terms of the composite
operators $u(x_{\perp})$. As is well known, this correlation
function is, for imaginary times, determined by the free (interaction)
energy $V$
of a static quark-antiquark pair at temperature $T=1/L$ (cf. Eq.~(\ref{SP24})),
and therefore allows  us to further study important properties of the
Z$_2$ symmetric
phase. Furthermore, this discussion will offer the possibility to display
within perturbation theory certain
characteristics of the modified axial gauge, in particular its high
momentum behaviour.

\subsection{ Polyakov Loop Correlator to Order $g^{2}$}

The novel aspects of the following development are related to
the ultralocality property of the Polyakov loops. As a consequence of
ultralocality, the Polyakov loop correlator is given by a one particle
{\em irreducible} 2-point function, i.e., the vacuum polarization  rather than by
the one particle {\em reducible} Green function of standard Gaussian variables.
This structural change implies a very different physics content of the
relation (\ref{SP24})
already at the perturbative level.
More precisely, according to the definition of the field $u$
(Eq.~(\ref{I54})),
the correlation function can be identified with
the color neutral, $\mu = \nu = 3$ component of the vacuum polarization
tensor $\Pi^{\mu \nu}$, evaluated with the external Polyakov loop vertices of
Eq.~(\ref{q4b}).
Thus, up to a constant, the interaction energy is related to the vacuum
polarization by
\begin{equation}
V(r)=-\frac{1}{L} \ln w(r)
\label{q13}
\end{equation}
with
\begin{equation}
w(r)= \int \frac{d^3 p}{(2\pi)^3}
\Pi_{\rm gg}^{33}(ip_0,p_1,p_2)e^{i \vec{p}\vec{r}} \ , \qquad
r \neq 0       \ .
\label{q14}
\end{equation}
In one loop approximation (Fig.~3) $\Pi^{33}_{\rm gg}$ is given  by
\begin{equation}
\Pi_{\rm gg}^{33}(p) = \frac{2 g^2}{L} \sum_{q_3} q_3^2
\int \frac{d^3q}{(2\pi)^3} \sum_{a,b \neq 3}D_{\mu \nu}^{ab}(q,
q_3) D^{ab, \mu \nu} (p-q,q_3)            \ .
\label{q15}
\end{equation}
After performing the sum over color and Lorentz indices, the divergent
3-dimensional integration is most conveniently carried out in dimensional
regularization with the result
\begin{eqnarray}
\label{q16}
\Pi_{\rm gg}^{33}(p)& = & \frac{ig^{2}}{4\pi L}\sum_{q_{3}}q_{3}^{2}
\Bigg[\left(\frac{p^{3}}{2(q_{3}^{2}+M^{2})^{2}}-\frac{2p}{q_{3}^{2}+
M^{2}}+\frac{4}{p}\right)\ln\left({\frac{2\sqrt{q_{3}^{2}+M^{2}}+p}{2
\sqrt{q_{3}^{2}+M^{2}}-p}}\right) \nonumber\\
& - & \frac{2p^{2}}{(q_{3}^{2}+M^{2})^{3/2}}  \Bigg] e^{-\lambda|q_3|}\ .
\end{eqnarray}
In this regularization procedure,  divergencies are encountered only in the
final sum over the ``Matsubara momenta'' $q_{3}$ which as indicated above we
regularize by a heat kernel method. The final expression has to be evaluated
numerically.
 For vanishing  momentum, Eq.~(\ref{q16}) simplifies to
\begin{equation}
  \label{q17}
\Pi_{\rm gg}^{33}(0) = \frac{ig^{2}}{\pi L}\sum_{q_{3}}\frac{q_{3}^{2}}
{\sqrt{q_{3}^{2}+M^{2}}} e^{-\lambda|q_3|} \ .
\end{equation}
The complete vacuum polarization associated with the
Polyakov loop variables is obtained by adding the (momentum independent)
tadpole contribution
\begin{equation}
\label{q18}
\Pi_{\rm tp}^{33} = \frac{ig^{2}}{\pi L}\sum_{q_{3}}
\sqrt{q_{3}^{2}+M^{2}}
e^{-\lambda|q_3|}
\end{equation}
which yields (for $p=0$)
\begin{equation}
\Pi_{\rm gg}^{33}(0)+\Pi_{\rm tp}^{33} = \frac{1}{3}\frac{i g^2}{ L^2}
\left(1+\delta \pi\right) \ ,
\label{q19}
\end{equation}
with
\begin{equation}
\label{q20}
\delta \pi = \frac{6L}{\pi} \sum_{q_{3}}\left[\frac{q_{3}^{2}+M^{2}/2}
{\sqrt{q_{3}^{2}+M^{2}}}-|q_{3}|\right] =  0.024
\end{equation}
characterizing the influence of the geometrical mass. As in Eq.~(\ref{I47})
the
unusual sign for the vacuum polarization is obtained. Unlike for massless
Gaussian variables, this sign does not imply a dynamical
instability for the  ultralocal Polyakov loops.

For large spacelike momenta, the vacuum polarization is given by
\begin{eqnarray}
\Pi_{\rm gg}^{33}(p) & \approx & -\frac{ig^2}{4\pi} \left[
(-p^2)^{3/2}L S_1
\right. \nonumber \\
& & \left.
-  \frac{p^2}{9 \pi} \left\{ 33 \left( \ln \frac{\lambda (-p^2)^{1/2}}
{2} + \gamma \right) + 1 \right\}
+ \frac{(-p^2)^{1/2}}{L} S_2 \right]
\label{q21}
\end{eqnarray}
with the sums
\begin{eqnarray}
S_1 & =  & -\frac{\pi}{2L^{2}}
\sum_{q_3}  \frac{q_3^2}{(q_3^2+M^2)^2}=\frac{-\pi}{16}\left(
\frac{\tanh{LM/2}}{LM/2}+\frac{1}{\cosh^{2} LM/2}\right)
=  -0.32
\nonumber \\
S_2 & =  &  2\pi M^2
\sum_{q_3} \frac{1}{q_3^2+M^2} = \pi L M \tanh{\frac{LM}{2}}
=  1.83
\label{q22}
\end{eqnarray}
In the modified axial or temporal gauge, the leading term of the vacuum
polarization at large momenta is not of the familiar form $p^{2}\ln{p^{2}}$.
Although such a  term is  present in the expansion, it is
subleading to the $L p^{3}$
contribution. The leading asymptotic term is generated by
the ``gauge term" ($\sim
p^{\mu} p^{\nu} /(p_3^2+M^2)$) in the propagator (\ref{q1}) as can be seen
from
the dependence of the expression (\ref{q15}) on size or inverse
temperature $L$
of the system. Summation over the wave numbers $q_{3}$ contributes a factor
$L^{-1}$, and so does each of the two  momenta from the vertices
($q_{3}^{2}$). A gauge term in the propagator contributes a factor
$L^{2}$. Thus the product of the two gauge
terms gives rise to the leading linear dependence on $L$ which in turn
implies the  $p^{3}$ dependence in Eq.~(\ref{q21}).

One might argue that this large momentum  behaviour of the vacuum polarization
is a gauge artefact. Indeed similar deviations from  standard large
momentum behavior are also present in QED. In Appendix A, we have sketched the
calculation of  the one loop fermion self energy at large
momenta in a modified axial gauge. The leading term,
\begin{equation}
\label{q23}
\Sigma (p) \approx  \frac{e^{2}}{192}/ \!\!\! p_{\perp} L
\sqrt{-p_{\perp}^{2}}                      \ ,
\end{equation}
is generated by the corresponding gauge term in the photon propagator.
In this Appendix, it is also shown that gauge terms do not contribute
to  gauge invariant quantities  such as
the two point function
\begin{displaymath}
\langle 0|T(\bar{\psi}_\beta(y)\,\exp \left\{
-i\,e\int\limits_x^y\!ds^\mu\,A_\mu\right\}
\,\psi_\alpha(x))|0\rangle
\end{displaymath}
obtained by insertion of  a gauge string into the definition of  the fermion
propagator. In QCD this alternative does
not exist. The Polyakov loop correlator is (before gauge fixing) a gauge
invariant quantity. Unlike in QED where, in the  process of gauge fixing,
the fermion field operators are  made gauge invariant by attaching gauge
strings which extend over the whole compact direction, the appearance of the
extension $L$ in the leading term of Eq.~(\ref{q22}) is due to the  intrinsic
property of the Polyakov loops of winding around the compact direction  and is
not generated by the gauge choice.

The complete
momentum dependence of the vacuum polarization (tadpole included) associated
with the Polyakov loops can be evaluated numerically with the help of
Eq.~(\ref{q16}) and Appendix B.
The physics content of the
Polyakov loop correlator is exhibited by the following discussion of the
static quark-antiquark interaction energy (Eqs. (\ref{q13}), (\ref{q14})).
Instead of performing directly the  Fourier
transform of $\Pi_{\rm gg}^{33}$ (cf. Eq.~(\ref{q14})), it is more convenient to
transform the Euclidean gluon propagator
to $r$-space,
\begin{equation}
w(r) = \frac{4 g^2}{L} \sum_{q_3}
q_3^2 \tilde{D}_{\mu \nu} (r,q_3)\tilde{D}_{\mu \nu} (r,q_3)
\label{q190}
\end{equation}
with
\begin{equation}
\tilde{D}_{\mu \nu}(r,q_3) = \left[ \delta_{\mu \nu} - \frac{1}{\omega_3^2}
\partial_{\mu} \partial_{\nu} \right] \frac{1}{4\pi r} e^{-\omega_3 r}
\label{q200}
\end{equation}
and
\begin{equation}
\omega_3 = \sqrt{q_3^2+M^2} \ .
\label{q210}
\end{equation}
Performing the differentiations and summations over Lorentz indices,
we obtain the sum
\begin{equation}
w(r)=\frac{g^2}{2 \pi^2 L r^6}
\sum_{q_3} \frac{q_3^2 e^{-2\omega_3 r}}{\omega_3^4} \left(
3+6(\omega_3 r) +5 (\omega_3 r)^2+ 2 (\omega_3 r)^3+ (\omega_3 r)^4
\right)
\label{q220}
\end{equation}
which can only be evaluated numerically for general $r$. Taking the
logarithm of
Eq.~(\ref{q220}), we see that the $r$-dependent part of the potential
does not contain $g^2$ anymore, i.e., it is purely ``kinematical"
at one loop order.
For small separations
a logarithmic potential is obtained
\begin{equation}
V(r) \approx \frac{6}{L} \ln (\mu r)+ \mbox{O}(1)
\ , \qquad r/L \ll 1
\label{q23a}
\end{equation}
($\mu$ is an arbitrary scale).
The leading $1/r^6$ term in this small distance expansion of $w$
is the Fourier transform of the leading term $L(-p^2)^{3/2}$ in
$\Pi_{\rm gg}^{33}$
for large spacelike momenta, and is therefore generated
from the gauge terms. This will be seen to be true
also in higher order. For large separations the potential of static charges
increases linearly with the separation. The dominant contribution
to the sum (\ref{q22}) comes from the lowest Matsubara momenta
($n=0, -1$) and we find
\begin{equation}
V(r) \approx  \sigma_{\rm pert} r + \frac{2}{L}
\ln (\mu r) + \mbox{O}(1)
\ , \qquad r/L \gg 1 \ .
\label{q24}
\end{equation}
Here the leading contribution is determined by the singularity of the
vacuum polarization $\Pi_{\rm gg}^{33}(p)$ closest to the real axis.
The exponential fall-off and therefore the perturbative ``string tension"
(cf. Eq.~(\ref{SP28})) is
determined by the  threshold energy  $2\omega_3|_{n=0}$  for producing
two charged gluons.
We also note that for infinite extension or zero temperature, the interaction
energy vanishes to this order. As in other evaluations of the interaction
energy, this  has to be interpreted as arising from the cancellation between
singlet and triplet contributions to the free energy. In this way the
free energy actually becomes the color averaged interaction energy which
vanishes to lowest order. This degeneracy does however not persist at finite
extension or temperature where for instance neutral and charged gluons are
distinguished by geometrical mass term and boundary conditions. This
distinction disappears for large $L$; however, it is
independent of the coupling
constant and thus gives rise to this ``kinematical'' one loop result. Finally,
we stress once more that due to the ultralocality of the Polyakov loop
variables, the ``wrong'' sign of  $\Pi^{33}(0)$ does not give rise to an
unphysical behaviour of the static quark interaction energy at large distances.

\subsection{Gluon Two-Point Function to One Loop}

In this section, we continue to develop perturbation theory in
the modified axial gauge. We will evaluate the gluon two-point function
to order $g^2$ and thereby address issues of  regularization and
renormalization. A topic  of special importance will be the asymptotic
behaviour of the gluon propagator at large  momenta.

For both the technical evaluation as well as for the physics
interpretation it is advantageous to separate the $g_{\mu \nu}$
and the $p_{\mu} p_{\nu}$ contributions to the gluon propagator $D_{\mu
\nu}^{ab}(p)$, Eqs. (\ref{q1}), (\ref{q3}).
We first evaluate the vacuum polarization insertion into the charged gluon
propagator, Fig.~4a. The formal expression
for this insertion is
\begin{equation}
\Pi_{\rm gg}^{\mu \nu} (p,p_3)=\frac{g^2}{L} \sum_{q_3}
\int \frac{d^3 q}{(2\pi)^3} \frac{N_0^{\mu \nu} + N_1^{\mu \nu}
+N_2^{\mu \nu} }{(q^2-q_3^2-M^2+i\epsilon)((q-p)^2-(q_3-p_3)^2+i\epsilon)}
\label{q54}
\end{equation}
The choice of the momenta is indicated in Fig.~4a. The loop momentum
$(q,q_3)$ denotes the charged gluon momentum. According
to the separation of the propagators into the $g_{\mu \nu}$ and the gauge
term (cf. Eqs. (\ref{q1}), (\ref{q3})), the vacuum polarization has been
broken up into contributions arising from none, one or two gauge terms
\begin{eqnarray}
N_{0}^{\mu \nu} & = & g^{\mu \nu} (2 q^2+5 p^2-2 qp)- 3 p^{\mu} p^{\nu} +
6 q^{\mu} q^{\nu} - 3 (p^{\mu} q^{\nu} + q^{\mu} p^{\nu} ) \ ,
\nonumber \\
N_{1}^{\mu \nu} & = & \frac{1-\delta_{q_3,p_3}}{(q_3-p_3)^2} \left[
-g^{\mu \nu} (q^2-p^2)^2+ p^{\mu} p^{\nu} (p^2-2 q^2) + q^{\mu}
q^{\nu} (q^2-2p^2) \right.
\nonumber \\
& & \left. + (p^{\mu}q^{\nu}+q^{\mu} p^{\nu} )(pq) \right]
+ \frac{1}{q_3^2+M^2} \left[ q \to p-q \right] \ ,
\nonumber \\
N_2^{\mu \nu} & = & \frac{1-\delta_{q_3,p_3}}{(q_3-p_3)^2(q_3^2+M^2)}
(p^2 q^{\mu}- qp\, p^{\mu})(p^2 q^{\nu} - qp\, p^{\nu} ) ,
\label{q55}
\end{eqnarray}
with their characteristic denominators. As above, the 3-dimensional
integrations are performed within dimensional regularization  and the
resulting divergent sums  are computed with heat-kernel regularization.
The 3-dimensional integrations can be carried out in closed form,
and the resulting expressions are given in Appendix B. The subsequent
sum over the 3-momenta cannot be performed analytically due to the gluon mass.

We first discuss the regularization dependent contributions
to the charged gluon self energy. The product of the two gauge terms
makes the momentum sum over $N_2^{\mu \nu}$ in Eqs. (\ref{q54}), (\ref{q55})
convergent,
while $N_{0,1}^{\mu \nu}$ give rise to divergent momentum sums.
The regulator dependent terms are
\begin{equation}
\Pi_{\rm gg}^{\mu \nu}(p,p_3)  =  i\frac{g^2}{\pi^2} \left\{
\frac{2}{3 \lambda^2} g^{\mu \nu} - \frac{11}{12} \ln
\lambda\left[ (p^2-p_3^2+\frac{2}{11}M^2)
g^{\mu \nu} - p^{\mu} p^{\nu} \right] \right\} + ...
\label{q56}
\end{equation}
In addition the tadpole contribution of  Fig.~4b has to be included.
Evaluation of
this contribution in the same scheme as above yields
\begin{equation}
\Pi_{\rm tp}^{\mu \nu} (p,p_3) = i \frac{g^2}{\pi^2}
\left( -\frac{2}{3\pi^2\lambda^2} + \frac{1}{6}M^2 \ln \lambda \right)
g^{\mu \nu} + ...
\label{q57}
\end{equation}
The quadratically divergent mass corrections appearing in the vacuum
polarization and tadpole cancel each other, as does the $M$-dependent
logarithmic divergence.
The remaining divergence,
\begin{equation}
\Pi^{\mu \nu}(p,p_3)= \Pi_{\rm gg}^{\mu \nu}(p,p_3) +
\Pi_{\rm tp}^{\mu \nu}(p,p_3)
= i \frac{g^2}{\pi^2}\left\{ -\frac{11}{12}
\ln  \lambda \left[ (p^2-p_3^2)g^{\mu \nu}-p^{\mu} p^{\nu} \right]
\right\}
+ ...
\label{q58}
\end{equation}
can be eliminated by wave function renormalization, i.e., by inclusion of a
counterterm to the Lagrangian of Eq.~(\ref{J6})
\begin{equation}
\label{q58a}
\delta {\cal L} =-(Z_{\Phi}-1)\left[ \frac{1}{2}
\left(\partial_{\mu}\Phi_{\nu}^{\dagger}-\partial_{\nu}\Phi_{\mu}^
{\dagger}\right) \left(\partial^{\mu}\Phi^{\nu}-\partial^{\nu}\Phi^
{\mu}\right)+ \partial_{3}\Phi_{\mu}^{\dagger}\partial^{3}\Phi^{\mu}\right] .
\end{equation}
The choice
\begin{equation}
Z_{\Phi} =  z_{\Phi}  - \frac{11 g^2}{12 \pi^2} \ln \lambda \
\label{q62}
\end{equation}
with  finite $z_{\Phi}$ ($ z_{\Phi} = 1 $ in  minimal subtraction)
gives rise to a finite charged
gluon propagator
which (suppressing trivial color labels) can be written as
\begin{equation}
\tilde{D}_{\mu \nu} \approx  D_{\mu \nu}+ i D_{\mu \sigma} \Pi^{\sigma \rho}
D_{\rho \nu}\approx
\frac{z_{\Phi}}{p^2-p_3^2-z_{\Phi} M^2+i \epsilon} \left[ -g_{\mu \nu} +
\frac{p_{\mu}
p_{\nu}}{p_3^2+z_{\Phi} M^2} \right] .
\label{q61}
\end{equation}
Thus the  divergent contributions to  vacuum
polarization and tadpole do not change  the gauge structure of the charged
gluon propagator.

The neutral gluon propagator can be discussed in the same way. The free
propagator (\ref{q3}) does not contain a (geometrical) mass term nor is such
a term
generated at one loop order.
Again only wave function renormalization is required.

The behaviour of the charged gluon self energy for large spacelike
momenta is of interest in particular for the calculation of the Polyakov
loop correlation function at small distances. As for the vacuum
polarization correction to the Polyakov loop correlator calculated above,
also here the leading contributions arise from the gauge terms of the gluon
propagators. It is convenient to decompose the charged gluon self-energy
as follows,
\begin{equation}
\Pi_{\rm gg}^{\mu \nu}(p,p_3) = i g^2 \left( \left[ g^{\mu \nu}
\frac{p^{\mu} p^{\nu}}{p^2} \right] \Pi_{\rm gg}^{(1)}(p,p_3)
+ g^{\mu \nu} \Pi_{\rm gg}^{(2)}(p,p_3) \right) \quad (\mu, \nu=0,1,2)
\ .
\label{q63}
\end{equation}
This decomposition makes use of the covariance in 2+1 dimensional Minkowski
space
which is respected by the 2+1 dimensional residual gauge fixing
as well as the regularization procedure. The leading terms in the
asymptotic expansion for large spacelike momenta $p^2 \to -\infty$ are
\begin{eqnarray}
\Pi_{\rm gg}^{(1)}(p,p_3) & \approx &
\frac{(-p^2)^{5/2}}{64 L} \sum_{q_3 \neq p_3}\frac{1}
{(q_3^2+M^2)(q_3-p_3)^2}  \ , \nonumber \\
\Pi_{\rm gg}^{(2)}(p,p_3) & \approx &
\frac{(-p^2)^{1/2}}{32 L} \left( 2 - \sum_{q_3\neq p_3}
\frac{(p_3^2-2 q_3 p_3 -M^2)^2}{(q_3^2+M^2)(q_3-p_3)^2}\right) \ .
\label{q64}
\end{eqnarray}
Higher order terms are given in Appendix B. We note the difference in the
leading power of $\Pi_{\rm gg}^{(1)}$ and $\Pi_{\rm gg}^{(2)}$. The
$(-p^2)^{5/2}$ behaviour results when retaining only the gauge terms in the
two gluon propagators and is therefore determined by $N_2^{\mu \nu}$
of Eq.~(\ref{q55}). The resulting $p_3^{-4}$ dependence
together with the $1/L$
associated with the momentum sum requires, for dimensional reasons, the fifth
power in $p$. Furthermore, using the transversality property of  the three
gluon vertex
\begin{equation}
V_{3,\lambda \mu \nu}(p,q,r,)p^{\lambda}q^{\mu}r^{\nu} = 0 \ ,
\label{q29}
\end{equation}
it is easily seen that this term must be transverse, i.e., vanish
after contraction with $p_{\mu}$ or $p_{\nu}$. In turn, the large momentum
behaviour of $\Pi_{\rm gg}^{(2)}$ is obtained
by retaining in one of the propagators the gauge term. By simple power
counting a $p^3$ behaviour is expected. However integration over
$N_2^{\mu \nu}$ in Eq.~(\ref{q54}) yields a vanishing coefficient.

\subsection{Polyakov Loop Correlator to Order $g^{4}$ at Large Momenta}

In this section, we shall calculate the asymptotic behavior of the Polyakov
loop correlator to order $g^4$ for large momenta. As our one loop results
indicate, this asymptotic behavior is determined by the gauge terms in the
gluon propagators
and so is the related free energy of a quark-antiquark system. On the other
hand, by asymptotic freedom, a Coulomb-like quark-antiquark interaction
energy is expected for sufficiently  small separations. It is thus of
conceptual interest how the expected $1/r$ behavior arises given the unusual
asymptotic form of the propagators.

Three of the four Feynman diagrams shown in Fig.~5 contribute to order
$g^{4}$ to the Polyakov loop correlator. Diagram 5d vanishes since the
external vertices are linear in the independent summation variables $k_3, q_3$
and otherwise only even powers of these variables arise.
Although different from zero, diagram 5c,
the tadpole insertion in the one loop diagram, is asymptotically of the
same order in $p$ as the one loop diagram and therefore subleading as
will be seen. We are thus left with the self-energy 5a and
vertex correction 5b. As illustrated above, the  leading large momentum
behavior
can be analysed by distinguishing between the $g^{\mu\nu}$ and gauge term
($p^{\mu}p^{\nu}$) contributions to the gluon propagators
(Eqs. (\ref{q1}), (\ref{q3})) and classifying accordingly the contributions to a
given diagram.
The highest conceivable power of $L$ of the diagrams 5a, 5b  arises
when retaining in all the 5 propagators the gauge terms ($\sim (L^2)^5$); both
double sum and external vertices contribute a factor $L^{-2}$ giving rise to
a $L^6p^8$ behaviour. However, the coefficient of this term vanishes
due to to the transversality of the three
gluon vertex, Eq.~(\ref{q29}).
Therefore  a $g_{\mu \nu}$  contribution to at least one of the
propagators has to be kept, avoiding thereby appearance of a vertex with three
adjacent gauge terms. In Fig.~6 the two contributions to the diagrams 5a, 5b
are shown with the marked propagator denoting the $g_{\mu \nu}$ contribution
(gauge terms otherwise). Superficially, these contributions give rise to
a $L^4 p^6$ dependence. Dimensionally regularized integration yields however
a vanishing coefficient and thus at large momenta, this contribution actually
behaves as $L^2 p^4\ln(-p^{2}) $.
For consistency, at this order, one also has to take into account all
the terms generated from the original diagrams 5a, 5b by retaining
$g_{\mu \nu}$ in two out of the 5 propagators, dropping the
$q_3$ dependence of the denominators. This yields already a fairly large
number of terms which are identified in Figs. 7 and 8. Several
simplifications
occur before the final evaluation: Diagram 7b vanishes in
dimensional regularization, diagram 7c has a lower power of
$L$ than naive counting due to a divergent $n$-sum and can be discarded,
and diagram
8a vanishes due to the $k_3, q_3$ sum. Thus for the following 6 distinct
subdiagrams 6a, 6b, 7a, 7d, 8b and 8c, the coefficients of the leading
$L^2 p^4\ln(-p^{2})$ terms have to be calculated. A final comment concerns the
appearance of the logarithmic momentum dependence. Integration over one of the
two loop momenta ($k$) generates odd powers of the second loop momentum
($(-q^{2})^{n/2})$ which in the subsequent dimensionally regularized
integration leads to poles in the $\Gamma$ function and thereby to the standard
logarithmic momentum dependence.

In a rather lengthy calculation, we have determined the
individual contributions of
the subdiagrams of Figs. 6 -- 8.
The coefficients of the (two loop) Polyakov loop correlation function
which multiply the leading asymptotic term
\begin{equation}
\frac{ig^4 L^2}{2 \pi^2} p^4  \ln (-p^2) \ ,
\label{q47}
\end{equation}
are listed in Table I
(diagram 7a gives zero).
Adding up all the terms and including  the result of the one loop
calculation (Eq.~(\ref{q21})), we finally obtain
\begin{equation}
\left. \Pi_{\rm gg}^{33}(p)\right|_{\rm 1-loop}+\left. \Pi_{\rm gg}^{33}(p)
\right|_{\rm 2-loop}  \approx
0.32 i\frac{g^2}{4\pi}L(-p^2)^{3/2}\left[1-0.043\frac{g^2}{4\pi}L
(-p^2)^{1 /2}\ln(-p^2)\right] .
\label{q48}
\end{equation}
\subsection{Interaction of Static Charges at Short Distances}
Fourier transformation of the above result yields the short distance behavior
of the Polyakov loop correlator
\begin{equation}
{\cal D}(r)  \approx   0.31 \,\frac{ig^2L}{\pi^2r^6}
\left(1+ 0.67 \,
\frac{g^2}{4\pi}\frac{L}{r} \right)
\label{q49}
\end{equation}
and thus, up to an additive constant, the free energy associated with static
charges at small separations
\begin{equation}
V(r)  \approx  \frac{6}{L} \ln (\mu r)  -0.67 \,
\frac{g^2}{4\pi}\frac{1}{r} \ .
\label{q50}
\end{equation}
We first note that the expansion parameter in this perturbative treatment of
the Polyakov loop correlator is $g^2 L/4 \pi r$ and not  $g^2/4 \pi$. This is
to be expected if the interaction energy contains a Coulomb-like contribution
$g^2 /4 \pi r$. In this case, like in QED, expansion of the exponential
$\exp(-L V(r))$
requires
\begin{equation}
\frac{g^2}{4\pi}\frac{L}{r} \ll 1 \ .
\label{q75}
\end{equation}
For sufficiently large extensions $L$ and by asymptotic freedom,
\begin{displaymath}
r \ll L \quad \mbox{and} \quad r\ll 1/\Lambda_{\rm QCD} \ ,
\end{displaymath}
we actually do expect the interaction energy to be given by
\begin{equation}
\label{q75a}
V(r) \sim c \frac{g^2}{4\pi r} + \frac{1}{L} v(r)+....
\end{equation}
up to $1/L$ corrections and terms of higher power in the coupling constant.
It is remarkable that the expansion in the parameter (\ref{q75}) is generated by
the gauge terms in the gluon propagators; we recognize here the intimate
connection between the ultralocality property of the Polyakov loops and the
presence of the gauge terms in the gluon propagators. While by ultralocality
the Polyakov loop correlator gets reduced to a correlator of the composite
field $u(x_{\perp})$ (cf. Eq.~(\ref{J4})), the higher powers of $L/r$ in the
composite field correlator can only be generated by the gauge terms. Our
calculation also shows that the appearance of the combination
$\frac{g^2}{4\pi}(\frac{L}{r})^{n}$ with $n=1$ is by
no means trivial. As shown
above, dimensional arguments suggest $n=3$ or $n=2$
as leading terms and only due to
specific properties of propagators and couplings, the corresponding
coefficients turn out to be zero. In summary, the structure of our final
result for the Polyakov loop correlator is in agreement with general
expectations and strongly supports the consistency of the approach.

The value of the constant multiplying the Coulomb interaction further
characterizes the dynamics in the Z$_2$ symmetric phase.  In the confined
phase, calculation of the interaction energy either via the Polyakov loop
correlator or via Wilson loop  are expected to yield the same result which
to lowest order and for short distances is given by the
singlet potential
\begin{equation}
V_0(r) = - \frac{3}{4} \frac{g^2}{4\pi} \frac{1}{r} \ .
\label{q52}
\end{equation}
On the other hand, treating Polyakov loop variables as Gaussian variables
and applying standard perturbation theory yields the
color averaged potential (\cite{NADK86}, \cite{NADK862}) with the weight given
by the statistical
factor,
\begin{equation}
V_{\rm av}(r) = \frac{1}{4} (V_0(r) + 3 V_1(r)) =  \mbox{O}(g^4/r^2) \ ,
\label{q53}
\end{equation}
i.e., to lowest order a vanishing interaction energy is obtained.
Our approximative interaction energy is significantly different from this
color averaged potential and rather close to the singlet potential. This is a
consequence of having completely resolved the Gauss law. In a perturbative
resolution,
one gluon exchange is attractive in the
singlet and repulsive in the triplet channel; the interaction effects
exactly cancel when taking the trace over the color spins of the static charges.
With the Gauss law fully resolved, all states and operators are gauge
invariant.
The concept of singlet or triplet states is not meaningful anymore. The
perturbative triplet states are made gauge invariant by appropriate gluon
admixtures. To estimate the effect of such gluon admixtures, we change the
weight in the average potential,
\begin{equation}
e^{-L\tilde{V}_{\rm av}(r)} = \left(e^{3Lg^{2}/16\pi r}+ 3 e^{-L\Delta E}
e^{-Lg^{2}/16\pi r }\right)/ \left(1+ 3 e^{-L\Delta E} \right) \ .
\label{q53a}
\end{equation}
$\Delta E$ is the energy necessary to produce  gluons which compensate for
the color of the static charges in the triplet state. If we
identify this energy
with the threshold of charged gluons
\begin{equation}
\Delta E \approx \frac{1}{L}\sqrt{\frac{4}{3}\pi^{2}-2} \ ,
\label{q53b}
\end{equation}
a Taylor expansion in terms of the coupling constant yields
\begin{equation}
\tilde{V}_{\rm av}(r) = -0.65 \,
\frac{g^2}{4\pi}\frac{1}{r} \   .
\label{q53c}
\end{equation}
The agreement of this simple estimate with the above calculation
suggests that in the Z$_2$ symmetric phase the sector of
states  originating  from color  carrying states is shifted towards higher
energies relative to the sector of states originating from perturbative
singlet
states. We thus arrive from this consideration of the interaction energy at
short distances to conclusions very similar to those obtained in the
discussion of the ``perturbative'' linear confinement. In both cases we find
that the phase  realizing the center symmetry  apparently contains the
seeds for non-perturbative phenomena such as the energetic suppression of
colored states and a linear rise of the interaction energy with increasing
separation. However, the energy scale connected with these phenomena is
determined geometrically by the extension of the system
rather than by non-perturbative dynamics.
\subsection{Interaction of Static Charges in the Presence of Light Quarks}
In this paragraph we shall discuss the effect of dynamical quarks on the
interaction energy of static charges at large distances. Our analysis will be
a perturbative one, i.e., we shall assume that the fermionic ground state is
that of free fermions which satisfy the boundary conditions (\ref{I23}). The
change from anti-periodic boundary conditions to these quasi-periodic ones
accounts for the interaction with the Aharonov--Bohm fluxes generated by the
ultralocal Polyakov loop variables and, as will be seen, is the crucial
agent for producing the string breaking mechanism.

Starting point of our perturbative calculation is Eq.~(\ref{SP25}) which yields
for the asymptotic behaviour of the interaction $V$
\begin{equation}
\label{CEE1}
\exp{\left\{-L V\left(r \right)\right\}} \approx  \left|
\langle \Omega|u\left(0\right)|\Omega\rangle  \right|^{2}
\end{equation}
if the order field $u$ develops a non-vanishing vacuum expectation
value. Disregarding the coupling between quarks and gluons, this vacuum
expectation value has to be generated by the quarks and according to the
definition of $u$ (Eq.~(\ref{I54})) is given by
\begin{eqnarray}
\langle \Omega|u\left(x_{\perp}\right)|\Omega\rangle & = & -L \langle \Omega|
\bar{\psi} \left(x\right)  \frac{\tau_{3}}{2}\gamma_{3}
\psi\left(x\right)|\Omega\rangle \\
& = & \frac{L}{2}\left[ \langle \Omega|\bar{\chi}_{\alpha} \left(x\right)
\gamma_{3}
\chi_{\alpha}\left(x\right)|\Omega\rangle\Big| _{\alpha =1/4}-\langle \Omega|
\bar{\chi}_{\alpha} \left(x\right)  \gamma_{3}
\chi_{\alpha}\left(x\right)|\Omega\rangle\Big| _{\alpha =-1/4}\right] \ .
\nonumber
\label{CEE2}
\end{eqnarray}
In eliminating the color dependence we have introduced  fermion fields
\begin{equation}
\label{CF2}
\chi_{\alpha}(x) = \sum_{\pm s}\frac{1}{\sqrt{L}}\sum_{k_{3}} \int
\frac{d^{2}k}{2\pi}\sqrt{\frac{m}{E_{k}}} \left[b(k,s)u(k,s)
e^{-ikx}+d^{\dagger}(k,s)v(k,s) e^{ikx}\right].
\end{equation}
which satisfy the boundary condition
\begin{equation}
\label{CEE3}
\chi_{\alpha}(x_{\perp}, x_{3}=L) = e^{2i\pi\alpha}\chi_{\alpha}(x_{\perp},
x_{3}=0) .
\end{equation}
The divergent expression for the vacuum expectation value
\begin{equation}
\label{CF5}
j_{3}(\alpha,m,L)=\langle \Omega|\bar{\chi}_{\alpha} \left(x\right)
\gamma_{3}
\chi_{\alpha}\left(x\right)|\Omega\rangle = \frac{1}{L}\sum_{k_{3}} \int
\frac{d^{2}k}{(2\pi)^{2}} \frac{2 k_{3}}{E_{k}}
\end{equation}
is evaluated by performing the integral over transverse momenta in dimensional
regularization (scale $\mu$, $\omega \rightarrow 1$)
\begin{equation}
j_{3}(\alpha,m,L)=\frac{2\mu^{2-2\omega}}{L}
\frac{\Gamma \left(\frac{1}{2}-\omega\right)
}{\sqrt{\pi} \left(4 \pi \right) ^{\omega}}
\sum_{k_{3}}k_{3}\left(m^2+k_{3}^2\right)^{\omega -\frac{1}{2}}.
\label{CF6}
\end{equation}
The sum over the momenta
\begin{equation}
\label{CF7}
k_{3} = \frac{2\pi}{L}(n+\alpha)
\end{equation}
can be carried out explicitly (cf. \cite{EORBZ94}) and yields the finite result
\begin{equation}
\label{CEE4}
j_{3}(\alpha,m,L)= \frac{2 m^{2}}{\pi^{2} L} \sum_{n=1}^{\infty}
\frac{\sin 2\pi
n\alpha}{n} K_{2}(nmL) \ .
\end{equation}
The vacuum expectation value of the current vanishes if $\alpha=0$ or
$\alpha=1/2$, i.e., for periodic or antiperiodic boundary conditions. Only then
does the spectrum of fermionic states contain parity doublets (with the
exception of the parity even zero mode
for $\alpha=0$). We also note that $j_{3}$ changes sign with $\alpha$.
Finally in the two limiting cases $mL \rightarrow 0,\infty$ the
following explicit expressions are obtained
\begin{equation}
j_{3}(\alpha,m,L)\approx \left\{ \begin{array}{cc}
\sqrt{2} \left(\frac{m}{\pi L}\right)^{3/2}e^{-mL}\sin 2\pi\alpha &
\quad mL \gg 1 \\
\frac{4\pi}{3 L^{3}} \left( \alpha -3\alpha^{2}+2\alpha^{3}\right) &
\quad mL \ll 1  \end{array} \right.
\label{CE5}
\end{equation}
Thus for asymptotic separations, the interaction energy of two static charges
reaches the constant value
\begin{equation}
\label{CE6}
V(r) \approx 2m \ ,
\end{equation}
up to corrections of O($1/L$).
It is remarkable that this correct form of the asymptotics of $V$ appears at
this perturbative level. In standard perturbation theory quark loops  yield as
in QED an Uehling type correction \cite{Uehling}
to the Coulomb interaction $\delta V \propto
g^{4} \exp{(-2mr)}/r$. Perturbation theory in the center symmetric phase on the
other hand yields the coupling constant independent result (\ref{CE6}). Thus the
mechanism of string breaking by pair formation is apparently present in the
center symmetric phase already at the perturbative level. The calculation
also displays the important role
of the modification of the boundary conditions, i.e., the role of the
Aharonov--Bohm
fluxes. The string breaking mechanism would not be present if dynamical quarks
satisfied standard
anti-periodic boundary conditions, nor would it arise if the boundary
conditions would not differentiate
between the two color states of the quarks
($\tau_{3}$ in the b.c. (\ref{I23})).
\section{Conclusions}
We have presented an analysis of SU(2) Yang Mills theory and QCD with the
focus on the
role of the center symmetry and the dynamics of the Polyakov loops. We have
carried out
these investigations with a choice of the geometry where one of the spatial
directions is
compact. Our gauge choice consists in eliminating the corresponding
spatial component
of the gauge
field up to the eigenvalues of the  Polyakov loops winding
around the
compact direction. This representation of QCD at finite extension is connected
by a rotation in the Euclidean   with the (modified) temporal gauge
representation of QCD at finite temperature. The gauge fixing procedure
leading to this  modified axial gauge representation has been carried out
explicitly and completely. No
perturbative elements are required in for eliminating
redundant variables
and, as a consequence, the resulting gauge fixed formulation is, in the
absence of quarks,
manifestly symmetric under center symmetry transformations. This is in
contradistinction to perturbative elimination of redundant variables or,
equivalently, perturbative treatments of gauge fixing terms and associated
Faddeev--Popov
determinants. Such approaches necessarily inherit the residual symmetries of
the underlying U(1)$^{N^{2}-1}$ rather than those of the SU($N$) gauge
theory and thereby generate  the weak coupling phase of an Abelian
theory with the dynamics of the Polyakov loops similar to
those of  QED. The correlation function of these variables yields, as is
well known,  Coulomb like interaction energies for static charges
and exhibits, in next order, the phenomenon of Debye screening. On the other
hand, such properties cannot be shared by a phase in which the center
symmetry is realized. In such a phase, the Polyakov loop expectation value
necessarily vanishes, signaling an infinite free energy of a single
static quark. Concomitantly the center symmetric phase does not screen color
charges; perturbative evaluation of the Polyakov loop correlator yields
imaginary masses for these degrees of freedom indicating instability of the
perturbative Polyakov loop vacuum. The physical vacuum has been shown to be
the vacuum of
ultralocal rather than Gaussian  degrees of freedom, i.e., degrees of freedom
which are essentially
inert and  can propagate only via their coupling to other degrees of
freedom.

This property of ultralocality of the Polyakov loop variables induces
significant changes in the formalism. Most importantly, it
effectively identifies propagator and  self-energy, i.e., the connected
with the one particle irreducible 2-point function. With this
structural change, confining interactions emerge as naturally from the
propagator
as do Yukawa or
Coulomb potentials in the case of Gaussian variables. This mechanism
is reminiscent of the emergence of confinement by formation of Gribov
horizons \cite{GRIB78}.
In both cases it is the limitation in phase space,
i.e., the finite
range of the functional integration, which is the source of confinement
phenomena. Essential differences arise by the gauge choice. In the
(modified) axial gauge, Gribov horizons appear for elementary degrees of
freedom while in Coulomb or Lorentz gauge, they restrict the phase space of
composite variables \cite{Lusch,Zwanz,Baal}.
Furthermore, with the identification of propagators and self energies,
simple relations could be established between  properties  of the confining
interaction between static quarks and the spectrum of gluonic states.
In particular a
non-vanishing value of the string constant in the thermodynamic limit has
been shown to  require the presence of an energy gap in the sector of
gluonic excitations which are coupled to
Polyakov loops. In the limit of infinite extension this gap must diverge.
Thereby this
sector of the Hilbert space gets effectively decoupled from low lying
excitations. A similar analysis has been  carried through for the Polyakov
loop
in the adjoint representation. The center symmetry does not force the vacuum
expectation value of the adjoint Polyakov loop to vanish and thus, depending
on the dynamics, a Coulomb or Yukawa potential acting between adjoint color
charges results. The behaviour under center symmetry
transformations constitutes the crucial difference between the states which
contribute to the correlation function of the Polyakov loops in either the
fundamental or the adjoint representation.

These results suggest the following
interpretation of the phases of the Yang--Mills theory. For sufficiently large
extension, the Hilbert space of physical states factorizes into the sector of
``low-lying'' states which are even under center symmetry transformations, and
the sector of odd states which is separated from the center symmetric
ground state by an energy
gap $\Delta E_{-} = \sigma L$ determined by extension and string
constant
and diverging in the thermodynamic limit. These two sectors are
completely disconnected. Furthermore, as follows from the absence of long
range strong interactions and as confirmed by lattice calculations, a gap
must be present also in the sector of even states; it is given by the
lowest glueball mass $\Delta E_{+} = M_{\rm g}$. With decreasing extension
(or equivalently
increasing temperature) the gap in the sector of odd states decreases till it
reaches a value $\Delta E_{-} \approx  \sigma T_{\rm c} \approx 1 \mbox{GeV}
\approx
\Delta E_{+} $. At this extension it apparently becomes energetically
favourable for the system to break the center symmetry, allowing thereby
for mixing of even and odd states. This in turn implies the vanishing of the
string tension. Due to the change in symmetry, this transition has to occur in
a discontinuous way.

Not only is the property of ultralocality of fundamental importance for the
structure of the center symmetric phase, it also is an essential ingredient
for establishing perturbation theory in the center symmetric phase.
After integrating out the Polyakov loops, expansion in the coupling constant
is possible without  ruining the center symmetry. As a consequence, certain
confinement
like properties can be obtained already within perturbation theory.
In particular a
linear rise in the interaction energy of static color charges is present
perturbatively. Quantitatively, the spectrum of gluonic excitations is poorly
described in perturbation theory and
does not yield a realistic value of the string constant. Perturbation
theory is however sufficient to reproduce the correct asymptotics of
the interaction energy if dynamical quarks  are present. Unlike ordinary
perturbation theory which yields a $1/r$
behavior modified  by a Uehling type potential arising from quark loops,
perturbation theory in the center
symmetric phase predicts correctly the interaction energy of
asymptotically  separated static color charges to be given by twice the quark
mass. In addition to these confinement like properties originating from
the Polyakov loop
dynamics, the perturbative center symmetric phase displays  non-trivial
properties of other gluonic degrees of freedom. In the process of elimination
of the Polyakov loops,  charged gluons, i.e., gluons
which are associated with the non-diagonal generators of the SU($N$) symmetry,
acquire a mass term and become coupled to space-time independent Aharonov--Bohm
fluxes. These modifications of the gluon dynamics are independent of the
coupling constant and geometrical, i.e., dependent on the extension. The
Aharonov--Bohm fluxes suppress Debye screening in the center-symmetric phase.
Presence of a mass term for charged
gluons and its absence for neutral gluons can be seen as a first hint
for Abelian
dominance of long range phenomena. It is remarkable that realization of the
center symmetry by proper gauge fixing  yields these characteristics of the
confined phase.

With the center symmetry realized, novel conceptual problems also arise in
the application of perturbation theory
to
short distance phenomena if confined variables are involved. This has been
illustrated in our discussion of the interaction energy of static charges
if calculated via the Polyakov loop correlator.
Clearly, irrespective of the realization of the center
symmetry, whenever the  separation of the charges $r$ is small,
i.e., $r\ll L=1/T$ and $1/r \gg \Lambda_{\rm QCD}$, this interaction energy must
be  given
by lowest order perturbation theory. However, perturbative evaluation of the
Polyakov loop correlator does not
reproduce the expected Coulomb-like behaviour, but rather seems to suggest a
$1/r^{2}$ dependence at short distances. However perturbation theory  not only
involves the  small coupling constant
$g^{2}/4\pi \ll 1$ but also the quantity
$g^{2}L/4\pi r =g^{2}/4\pi Tr$ which becomes large at short distances and
thereby turns the calculation of the Polyakov loop correlator into a
strong coupling problem. The Polyakov loop self-energies
are dominated at large momenta not by the familiar $p^{2}$ term with its
standard
logarithmic corrections but rather by extension or temperature dependent
$(g^{2}pL)^{n} p^{2}$ corrections. These unusual contributions to the
self-energy originate from the ``gauge terms'' of the gluon propagators. Their
presence is necessary for generating the proper short distance behavior, as is
the ultralocality of the Polyakov loops.

Our investigations represent a first, exploratory analysis  of the center
symmetric phase of
QCD. The focus has been on the consequences of the realization of the center
symmetry and in particular on those  properties which are reminiscent of the
confining phase of QCD and which emerge already at the perturbative
level. Nevertheless the wealth of
non-perturbative phenomena cannot be accounted for at this perturbative level.
Only  Polyakov
loop variables exhibit phenomena associated with confinement. Although other
degrees of freedom are significantly affected by the realization of the center
symmetry, confinement of these degrees of freedom is not manifest,
nor does it seem to be within the reach of perturbation theory. Self couplings
of the gluonic degrees of freedom may generate these non-perturbative
dynamics. We have however not been able to identify specific
mechanisms. Alternatively, it
might be necessary to extend the formalism by inclusion of singular gauge
field configurations when integrating out the Polyakov loop variables. Our
implicit restriction to smooth gauge fields may not be justified. When
diagonalizing the Polyakov loops, ambiguities arise whenever
the Polyakov loop passes through the center of the group.
These ambiguities yield the monopole like singularities characteristic for
Abelian projected QCD. Condensation of these monopoles could then lead, via the
dual Meissner effect, to confinement of  gluonic degrees of freedom also where
center symmetry does not dictate it. We are in the process of
studying the necessary modifications of the formalism.

Extension of our investigation to SU(3) or more generally SU($N$) Yang Mills
theory is of interest. As indicated in our work, the formalism can easily be
extended to higher groups. The construction of the effective Lagrangian
in its Abelian projected form is straightforward but technically more involved
(most of the necessary modifications have been worked out for SU(3) color
in \cite{Shifman}). The structure of the center symmetric phase is, in
general, more complex than for SU(2) color.
In particular, the Z$_3$ center symmetry of SU(3) Yang Mills theory
allows for a confinement-deconfinement transition which is first order
and may lead to formation of domains. Existence of such domains, the
Z$_N$ bubbles, is a controversial issue (cf. \cite{SMIL94}, \cite{BHAT92})
and our formalism may provide new perspectives. Unlike at finite
temperature, the center symmetry is an ordinary symmetry at finite
extension. It is canonically described by an operator which commutes
with the Hamiltonian and
thus formation of Z$_N$ bubbles under compression of the system is
conceptually simpler than bubble formation when heating the
system. Furthermore, application of our techniques to the large $N$-limit
might provide the possibility to study the weak coupling, confining phase
which e.g. has been invoked to connect large $N$-QCD with
string theory \cite{POLC92}.

Our final remarks concern the description of the deconfined phase. With the
center symmetry realized, the transition to the deconfined phase becomes, as it
should be, a dynamical issue. From this point of view, the standard procedure
to reach the deconfined phase by perturbative gauge fixing or
equivalently  by perturbative resolution of
the Gauss law appears to be rather arbitrary. Perturbative gauge
fixing prevents the system from reaching the confined phase at the
expense of not only breaking the center symmetry but also violating other
constraints imposed by local gauge invariance. Problems associated with such a
procedure appear explicitly in the standard derivation of Debye screening in
temporal or modified
temporal gauge. Naive treatment of the Polyakov
loop variables as Gaussian variables is not legitimate at any temperature.
Also in the
deconfined phase, these degrees of freedom have, as angular variables,
a finite range of definition
and remain ultralocal. On physical grounds, however, we
expect Debye screening to be a reasonable approximate concept at
sufficiently high temperatures. As
in the Debye--H\"{u}ckel theory where, in conflict with the Gauss law, one shows
screening of a test charge when introduced into a neutral system, in QCD one
apparently might have to
allow similarly for certain violations of the Gauss law rather than to insist
on it's exact implementation. However the non-linearities of the Gauss law
may invalidate such an approximation in QCD (inert background charges which
compensate for violations of the Gauss law do not exist) and lead to
inconsistencies in higher order. The divergence of finite temperature QCD
perturbation theory beyond $g^{6}$ \cite{LIND80} indicates the failure of
such an approach and, in agreement with results from lattice QCD
(cf. \cite{BEKL96}), suggests the
deconfined phase and its screening properties not to be accessible by standard
perturbation theory
(\cite{BRNI95}, \cite{BRAA95}). Although perturbative investigations are
not sufficient to describe the dynamics of the confinement-deconfinement
phase transition, they nevertheless are useful in identifying properties
which necessarily change when the center symmetry gets spontaneously broken.
These can be used in conjunction with stability requirements to
characterize the high temperature deconfined phase \cite{EKLT97}.

\vskip 1.0cm
We thank V.L. Eletskii and A.C. Kalloniatis for valuable discussions.
This work has been supported by the Bundesministerium f\"ur Bildung,
Wissenschaft, Forschung und Technologie.

\setcounter{equation}{0}
\renewcommand{\theequation}{A.\arabic{equation}}

\newpage

\section*{Appendix A: Electron Self-Energy in Axial Gauge}

In this appendix we show in lowest order perturbation theory that the
electron self energy in axial gauge is dominated at
large (transverse) momenta by the gauge terms in the propagator. The photon
propagator in axial gauge QED coincides for $p_{3} \neq 0 $ with the neutral
gluon propagator of
SU(2)-QCD (cf. Eq.~(\ref{q3}))
\begin{equation}
D_{\mu \nu}(p,p_3) = \frac{1}{p^2-p_3^2+i\epsilon}
\left[-g_{\mu \nu} + p_{\mu} p_{\nu}(1-\delta_{p_3,0})\frac{1}{p_3^2}
+p_{\mu} p_{\nu}\delta_{p_3,0}(1-\xi)\frac{1}{p^2+i\epsilon} \right]
\label{ESE1}
\end{equation}
with
\begin{equation}
p_3= \frac{2\pi n}{L} \ ,
\label{ESE2}
\end{equation}
but unlike QCD contains a 33-component describing photons polarized in the 3
direction and propagating in transverse space,
\begin{equation}
D_{33}(p) = \frac{\delta_{p_3,0}}{p^2+i\epsilon} \ .
\label{ESE3}
\end{equation}
We consider the  contribution to the self-energy which is generated
by the second term  in the  propagator (\ref{ESE1}),
\begin{equation}
\label{ESE4}
i\delta \Sigma (p) =
e^2 \frac{1}{L}\sum_{q_{3}}\int\!\frac{d^3\!q}{(2\pi)^3}
\frac{(1-\delta_{q_3,0})}{q^2-q_3^2+i\epsilon}\frac{q_{\mu} q_{\nu}}{q_3^2}
\frac{\gamma^\mu\,[(/ \!\!\! p  - / \hspace{-0.2cm} q) + m]\,\gamma^\nu}
{(p-q)^2-(p_{3}-q_{3})^2 - m^2 + i\epsilon} \ .
\end{equation}
Using the techniques developed in the main section, the leading term for large
transverse momenta can  straightforwardly be determined with the result
\begin{equation}
\label{ESE4a}
\delta \Sigma (p) \approx  \frac{e^{2}}{192}/ \!\!\! p_{\perp} L
\sqrt{-p_{\perp}^{2}}  \ .
\end{equation}
Thus in axial gauge at  large transverse momenta, the first order correction
exceeds the free self-energy by a power of $p_{\perp} L$.

Such an   unusual large $p$ asymptotics which, in the electron self energy, is
generated by the gauge terms of the photon propagator, is absent if we
start (before gauge fixing) from a gauge invariant quantity such as the
electron two point function with an appropriate gauge string insertion,
\begin{equation}
\label{ESE5}
\tilde{S}(x,y)_{\alpha\beta}  =
i \langle 0|T(\bar{\psi}_\beta(y)\,
\exp\left\{-i\,e\int\limits_x^y\!ds^\mu\,A_\mu\right\}
\,\psi_\alpha(x))|0\rangle \ .
\end{equation}
To order  $e^{2}$, this two point function is given by
\begin{eqnarray}
\label{ESE6}
\tilde{S}(x,y) & = &
S_{\rm F}(x,y)\left (
1 - i\,\frac{e^2}{2}\int\limits_0^1\frac{dx^\mu}{ds}
\int\limits_0^1\frac{dx^\nu}{ds'}\,D_{\mu\nu}(x(s),x(s'))
\right )\nonumber \\& &
{}+ e^2\int\limits_0^1\frac{dx^\mu}{ds}\int\!d^4\!z\,D_{\mu\lambda}(x(s),z)\,
S_{\rm F}(x,z)\,\gamma^\lambda\,S_{\rm F}(z,y)\nonumber \\& &
{}+i \frac{e^2}{2}\int\!d^4\!z\,d^4\!z'\,D_{\mu\lambda}(z,z')
[S_{\rm F}(x,z')\,\gamma^\nu\,S_{\rm F}(z',z)\,\gamma^\mu\,S_{\rm F}(z,y)
\nonumber \\
& &\hfill {}+ S_{\rm F}(x,z)\,\gamma^\mu\,S_{\rm F}(z,z')\,\gamma^\nu\,
S_{\rm F}(z',y)]
\end{eqnarray}
with $x^{\mu}(s)$ parametrizing the gauge string.
We split the photon propagator
\begin{equation}
\label{ESE7}
D_{\mu\nu}(q)=D_{\mu\nu}^{\rm F}(q)+D_{\mu\nu}^{\rm ax}(q)\quad  \mu,\nu = 0..3
\end{equation}
into the Feynman gauge propagator
\begin{equation}
\label{ESE8}
D_{\mu\nu}^{\rm F}(q) = \frac{-g_{\mu\nu}}{q^2+i\epsilon}
\end{equation}
and the contributions from the axial gauge terms,
\begin{equation}
\label{ESE9}
D_{\mu\nu}^{\rm ax}(q) = q_\mu\,d_\nu+q_\nu\,d_{\mu} \ ,
\end{equation}
with
\begin{equation}
\label{ESE10}
d_\mu(q) =
(1 - \delta_{q_3,0})\,\frac{1}{2q^2\,q_3^2}\,
q_\mu\left(1 - \delta_{\mu,3}\right)
+ \delta_{q_3,0}\,\frac{q_\mu}{(q^2 + i\epsilon)^2}\,(1 - \xi) \ .
\end{equation}
To this order in perturbation theory, $\tilde{S}$ (Eq.~(\ref{ESE6})) depends
linearly on the
photon propagator and therefore can be decomposed into  contributions
from the Feynman gauge photon propagator and  axial gauge terms, respectively,
\begin{equation}
\label{ESE11}
\tilde{S}(x,y) = S_{\rm F}(x,z) + \tilde{S}^{\rm F}(x,y)+
\tilde{S}^{\rm ax}(x,y) \ .
\end{equation}
Denoting by $\tilde{s}^{i}$ the contributions
to $\tilde{S}^{\rm ax}(x,y)$  with $i$
(=0,1,2) gauge strings,
\begin{equation}
\label{ESE12}
\tilde{S}^{\rm ax}(x,y)= \sum _{i=0,1,2}\tilde{s}^{i}(x,y) \ ,
\end{equation}
we find
\begin{equation}
\label{ESE13}
\tilde{s}^{0}(x,y) =  i e^2\int\!d^4\!z\,
S_{\rm F}(x,z)\,\gamma^\mu\,S_{\rm F}(z,y)\,
[d_\mu(x-z) + d_\mu(z-y)]
\end{equation}
\begin{eqnarray}
\label{ESE14}
\tilde{s}^{1}(x,y) & = &
i e^2\int\!(d_\nu(y-z) - d_\nu(x-z))\,d^4\!z\,S_{\rm F}(x,z)\,
\gamma^\nu\,S_{\rm F}(x,y)\nonumber \\
& & - e^2\int\limits_0^1\frac{dx^\mu}{ds}\,ds\,
                (d_\mu(x(s)-y) - d_\mu(x(s)-x))\,S_{\rm F}(x,y)
\end{eqnarray}
\begin{equation}
\label{ESE15}
\tilde{s}^{2}(x,y)=-ie^2\int\limits_0^1\frac{dx^\nu}{ds}\,(d_\nu(y-x(s)) +
d_\nu(x(s) - x))
\end{equation}
and thus
\begin{equation}
\label{ESE16}
\tilde{S}^{\rm ax}(x,y) = 0 \ .
\end{equation}

In the above equations, $d_{\mu}(x)$ denotes the Fourier transformed gauge
terms (\ref{ESE10}). In the derivation, we
have applied the  equation of motion of $S_{\rm F}$ and thereby made use of the
continuity equation. Our result implies the  independence of the large $p$
asymptotics of $\tilde{S}(x,y)$ of the extension $L$, provided of course
$L$ is not reintroduced
by the choice of the gauge string (e.g. if $x^{\mu}(s)$ is a path winding
around the compact direction).

\newpage

\setcounter{equation}{0}
\renewcommand{\theequation}{B.\arabic{equation}}

\section*{Appendix B: One Loop Gluon Self Energy for SU(2)}

Of interest for us are the self energies of neutral, ``electric"
($a_3$) and charged, ``magnetic" ($A_{0,1,2}$) gluons, which will
be given in all detail.
The results for neutral, ``magnetic" gluons can be inferred from
the charged ones as indicated below.

\subsection*{B.1 Gluon Loop, Exact Results}

Definition:
\begin{eqnarray}
\Pi^{\mu \nu}_{\rm gg}(p,p_3) & = &  \frac{i g^2}{4\pi L}
\left( \left[g^{\mu \nu}-
\frac{p^{\mu} p^{\nu}}{p^2}\right] \Pi_{\rm gg}^{(1)}(p,p_3)
+ g^{\mu \nu} \Pi_{\rm gg}^{(2)}(p,p_3)
\right)\qquad
(\mu, \nu = 0,1,2)
\nonumber \\
\Pi^{33}_{\rm gg}(p) & = & \frac{i g^2}{4\pi L} g^{33} \Pi^{(3)}_{\rm gg}(p)
\label{n1}
\end{eqnarray}
Parametrization:
\begin{eqnarray}
\Pi_{\rm gg}^{(i)} & = & \sum\,' \left\{ A_i \ln \left( \frac{x+y+P}
{x+y-P} \right) + B_i \right\}e^{-\lambda x_0} + C_i \qquad (i=1,2)
\nonumber \\
\Pi^{(3)}_{\rm gg} & = & \sum \left\{ A_3 \ln \left( \frac{2x+P}{2x-P}
\right) + B_3 \right\} e^{-\lambda x_0}
\label{n2}
\end{eqnarray}
We use the following notation:
\begin{eqnarray}
P & = & \sqrt{p^2}
\nonumber \\
x & = & \sqrt{q_3^2+M^2}
\nonumber \\
x_0 & = & |q_3|
\nonumber \\
y & = & |q_3-p_3|
\nonumber \\
z & = & \sqrt{p_3^2+M^2}
\nonumber \\
q_3 & = & \frac{2 \pi}{L} \left(n + \frac{1}{2}\right)
\label{n9}
\end{eqnarray}
The summation with a prime runs over all $n$
such that $q_3 \neq p_3$. A heat kernel
regularization is used.
The sums cannot be
performed analytically in general. The value for the charged gluon mass is
\begin{equation}
M^2 =  \left( \frac{\pi^2}{3}-2 \right) \frac{1}{L^2}
\label{n9a}
\end{equation}
Results for the coefficients $A_i, B_i, C_i$:
\begin{eqnarray}
A_1 & = & -\frac{1}{16} \left\{ \frac{P^5}{x^2 y^2}
+ 4 P^3  \left( \frac{1}{x^2} + \frac{1}{y^2} \right)
-8 P \left( \frac{x^2}{y^2}+\frac{y^2}{x^2} + 3 \right)
\right.
\nonumber \\
& & \left.
- \frac{16}{P} \left( x^2+y^2 \right)
+ \frac{3}{P^3}\left( \frac{x^6}{y^2} + \frac{y^6}{x^2}
+ 4 x^4 + 4 y^4 - 10 x^2 y^2 \right) \right\}
\label{n3}
\\
B_1 & = & \frac{(x+y)}{8}  \left\{ \frac{P^4}{x^2 y^2} -
P^2 \left( \frac{1}{x^2} + \frac{1}{y^2} - \frac{6}{x y} \right)
+ \left( \frac{x^2}{y^2}+ \frac{y^2}{x^2} -  \frac{4 x}{y}
- \frac{4 y}{x} - 2 \right) \right.
\nonumber \\
& & \left. + \frac{3}{P^2}\left( \frac{x^4}{y^2} + \frac{y^4}{x^2}
-  \frac{2 x^3}{y}-  \frac{2 y^3}{x} + 7 x^2 + 7 y^2 - 12 x y \right) \right\}
\label{n4}
\\
C_1 & = & - \frac{1}{8} \left(\frac{3 P^3}{z^2}
-13 P -\frac{7 z^2}{P} + \frac{9 z^4}{P^3}
\right) \ln \left( \frac{z+P}{z-P} \right)
\nonumber \\
& &
+ \frac{1}{4} \left( \frac{3P^2}{z} - 4 z + \frac{9 z^3}
{P^2} \right)
\label{n5}
\\
A_2 & = & \frac{\left( y^2-x^2 \right)^2 }{8}  \left\{
\frac{P}{x^2 y^2}- \frac{2}{P}
\left( \frac{1}{x^2} + \frac{1}{y^2} \right)
+ \frac{1}{P^3}
\left( \frac{x^2}{y^2}+\frac{y^2}{x^2} + 6 \right)
\right\}
\label{n6}
\\
B_2 & = & - \frac{(x+y)}{12}  \left\{
\frac{3 x^2}{y^2}+  \frac{3 y^2}{x^2}  - 22
\right.
\nonumber \\
& & \left.
+ \frac{3}{P^2}\left( \frac{x^4}{y^2} + \frac{y^4}{x^2}
\frac{2 x^3}{y}-  \frac{2 y^3}{x} + 7 x^2 + 7 y^2 - 12 x y \right) \right\}
\label{n7}
\\
C_2 & = & - \frac{1}{4} \left(
P -\frac{2 z^2}{P} - \frac{3 z^4}{P^3}
\right) \ln \left( \frac{z+P}{z-P} \right)
+ \frac{1}{6} \left( 11 z - \frac{9 z^3}
{P^2} \right)
\label{n8}
\\
A_3 & = & - \frac{x_0^2}{2}\left( \frac{P^3}{x^4}-\frac{4P}
{x^2}+\frac{8}{P} \right)
\label{n8a}
\\
B_3 & = & 2x_0^2 \frac{P^2}{x^3}
\label{n8b}
\end{eqnarray}
For neutral, magnetic gluons, $\Pi_{\rm gg}^{\mu \nu}$ can be taken over
with the following modifications: No summation restriction in Eq.~(\ref{n2}),
interpret $y$ as
\begin{equation}
y = \sqrt{(q_3-p_3)^2+M^2} \ ,
\label{n8c}
\end{equation}
and put all $C_i=0$.

\subsection*{B.2 Tadpole}

Definition:
\begin{eqnarray}
\Pi^{\mu \nu}_{\rm tp}(p,p_3) & = & \frac{i g^2}{4\pi L} g^{\mu \nu}
\Pi_{\rm tp}^{(\bot)}
\qquad
(\mu, \nu = 0,1,2)
\\ \nonumber
\Pi_{\rm tp}^{33} (p) & = & \frac{ig^2}{4\pi L} g^{33} \Pi_{\rm tp}^{(3)}
\label{n10}
\end{eqnarray}
Parametrization:
\begin{eqnarray}
\Pi_{\rm tp}^{(\bot)} & = & - \frac{4}{3} \sum (x+y) e^{-\lambda
 x_0}
\label{n11}
\\
\Pi_{\rm tp}^{(3)} & = & - 4 \sum x e^{-\lambda x_0}
\label{n11a}
\end{eqnarray}
Result (\ref{n11}) is valid for neutral magnetic gluons, provided
we interpret $y$ according to (\ref{n8c}).

\subsection*{B.3 Isolating Divergencies in the Sums}
To get convergent sums, add and subtract the following
sums derived from the large $x$ behaviour of the above results:
\begin{eqnarray}
\Delta \Pi_{\rm gg}^{(1)} & = & \frac{11}{3}P^2 \sum \frac{1}{x_0}
\nonumber \\
\Delta \Pi_{\rm gg}^{(2)} & = & \left( - \frac{11}{3}p_3^2
+ \frac{2}{3}M^2 \right) \sum \frac{1}{x_0} + \frac{8}{3} \sum x_0
\nonumber \\
\Delta \Pi_{\rm gg}^{(3)} & = & \left( \frac{11}{3}P^2
+ 2 M^2 \right) \sum \frac{1}{x_0} - 4 \sum x_0
\nonumber \\
\Delta \Pi_{\rm tp}^{(\bot)} & = &
- \frac{2}{3}M^2 \sum \frac{1}{x_0} - \frac{8}{3} \sum x_0
\nonumber \\
\Delta \Pi_{\rm tp}^{(3)} & = &
- 2 M^2 \sum \frac{1}{x_0} - 4 \sum x_0
\label{n11b}
\end{eqnarray}
Notice that
\begin{equation}
\Delta \Pi_{\rm gg}^{(2)} + \Delta \Pi_{\rm tp}^{(\bot)}  =
- \frac{11}{3} p_3^2 \sum \frac{1}{x_0} \ ,
\label{n11c}
\end{equation}
i.e., the quadratic divergence and the $M$-dependence of the logarithmic
divergence are cancelled. In $\Pi^{(3)}$, the
quadratic divergence persists,
\begin{equation}
\Delta \Pi_{\rm gg}^{(3)} + \Delta \Pi_{\rm tp}^{(3)}  =
\frac{11}{3} P^2 \sum \frac{1}{x_0} -8 \sum x_0 \ .
\label{n11d}
\end{equation}
However, this does not enter in our framework anyway.
The relevant heat-kernel regularized sums are
\begin{eqnarray}
\sum \frac{1}{x_0}e^{-\lambda x_0} & = & \frac{L}{\pi} \ln \left( \frac{2L}
{\pi \lambda} \right) \nonumber \\
\sum x_0 e^{-\lambda x_0} & = & \frac{L}{\pi \lambda^2}+ \frac{\pi}{6 L}
\label{n11e}
\end{eqnarray}
For neutral magnetic gluons, the results are the same as for charged ones.

\subsection*{B.4 Asymptotics for Large, Spacelike Momenta}

We consider the full self energy (gluon loop and tadpole) with the following
definitions, cf. Eqs. (\ref{n1}), (\ref{n10}),
\begin{eqnarray}
\Pi^{\mu \nu} & = & \Pi_{\rm gg}^{\mu \nu} + \Pi_{\rm tp}^{\mu \nu}
\nonumber \\
\Pi^{(1)} & = & \Pi_{\rm gg}^{(1)}
\nonumber \\
\Pi^{(2)} & = & \Pi_{\rm gg}^{(2)}+\Pi_{\rm tp}^{(\bot)}
\nonumber \\
\Pi^{(3)} & = & \Pi_{\rm gg}^{(3)}+\Pi_{\rm tp}^{(3)}
\label{n14}
\end{eqnarray}
For spacelike momenta, we introduce
\begin{equation}
P=i\bar{P} \ .
\label{n15}
\end{equation}
The gluon mass prevents us from deriving closed analytical results.
Therefore we split the calculation up as follows:
\begin{equation}
\Pi^{(i)}[M] = \Pi^{(i)}[0]  + \left( \Pi^{(i)}[M]-\Pi^{(i)}[0] \right)
\label{n15a}
\end{equation}
For the $M=0$ part, we get the following results where
terms of order less than $\bar{P}$ are only retained if they involve the
cut-off parameter,
\begin{eqnarray}
\Pi^{(1)}[0] & \approx &    \frac{\pi \bar{P}^5}{48 p_3^4}
\left(  \left( p_3 L_3 \right)^2 -9\right)
-  \frac{\pi \bar{P}^3}{24 p_3^2} \left( 3 + 2 \left( p_3 L_3 \right)^2
\right)
\nonumber \\
& &
+ \frac{L_3 \bar{P}^2}{9 \pi} \left\{ - 4 + 2 \pi  p_3 L_3
+ 33 \left( \ln \frac{\lambda \bar{P}}{2} + \gamma \right) + 12 \ln
\frac{\bar{P} L_3}{2\pi} \right.
\label{n16}
\\
& &
\left. - 12 \left[ \psi \left( \frac{p_3 L_3}{\pi} \right)
+ \frac{p_3 L_3}{\pi} \psi' \left( \frac{p_3 L_3}{\pi} \right)
\right] \right\}
+  \frac{\pi \bar{P}}{24} \left( 9 - 4 \left( p_3 L_3 \right)^2 \right)
+ \mbox{O}(\ln \bar{P})
\nonumber
\\
\Pi^{(2)}[0] & \approx &  \frac{\pi \bar{P}}{24}
\left( 3 - \left( p_3 L_3 \right)^2 \right)
+ \frac{11 L_3 p_3^2}{3\pi} \ln \lambda + \mbox{O}(\ln \bar{P})
\nonumber
\\
\Pi^{(3)}[0] & \approx & -\frac{\pi L_3^2 \bar{P}^3}{8}
+ \frac{L_3 \bar{P}^2}{9 \pi} \left\{ 33 \left( \ln \frac{\lambda \bar{P}}{2}
+ \gamma \right)+1 \right\} - \frac{8 L_3}{\pi \lambda^2}
+ \mbox{O} (1)
\label{n17}
\end{eqnarray}
The differences
\begin{equation}
\Pi^{(i)}[M]-\Pi^{(i)}[0] = \sum_n {\cal S}_n^{(i)} \bar{P}^n
\label{n18}
\end{equation}
have to be evaluated numerically.
The ${\cal S}_n^{(i)}$ are given by
\begin{eqnarray}
{\cal S}_1^{(1)} & = &  - \frac{\pi}{2} \sum\,'
\left\{ \frac{1}{y^2}\left(x^2-x_0^2 \right) + y^2 \left( \frac{1}{x^2}
-\frac{1}{x_0^2} \right) \right\}
\nonumber \\
{\cal S}_2^{(1)} & = & \frac{2}{3} \sum\,'
\left\{ \frac{1}{y^2}\left(x-x_0 \right) + y \left( \frac{1}{x^2}
-\frac{1}{x_0^2} \right) \right\}
\nonumber \\
{\cal S}_3^{(1)} & = & -\frac{3 \pi}{8} \left(\frac{1}{z^2}-\frac{1}{z_0^2}
\right) - \frac{\pi}{4} \sum\,'
\left( \frac{1}{x^2}
-\frac{1}{x_0^2} \right)
\nonumber \\
{\cal S}_5^{(1)} & = &   \frac{\pi}{16} \sum\,' \frac{1}{y^2}
\left( \frac{1}{x^2}
-\frac{1}{x_0^2} \right)
\nonumber \\
{\cal S}_1^{(2)} & = &   \frac{1}{4} {\cal S}_1^{(1)}
\nonumber \\
{\cal S}_1^{(3)} & = &  - 2\pi \sum x_0^2
\left( \frac{1}{x^2}
-\frac{1}{x_0^2} \right)
\nonumber \\
{\cal S}_3^{(3)} & = &  -\frac{\pi}{2} \sum x_0^2
\left( \frac{1}{x^4}
-\frac{1}{x_0^4} \right)
\label{n19}
\end{eqnarray}
All ${\cal S}_n^{(i)}$ for $n \geq 1$ not listed here are zero.

\newpage
\bibliographystyle{unsrt}

\newpage
\begin{center}
\vskip 1.0cm
\section*{Table I:  Contributions of 2-loop subdiagrams}

\begin{tabular}{|c|r|}\hline
diagram  &  value \hspace{0.6cm}  \\
\hline
6a  &  -0.67555E-03 \\
\hline
6b   &   -0.10516E-02  \\
\hline
7d &   -0.15106E-03 \\
\hline
8b &   0.14243E-03  \\
\hline
8c &   0.12037E-04 \\
\hline
\end{tabular}

\end{center}

\newpage
\section*{Figure Captions}
\vskip 1.0cm

\begin{itemize}

\item[{\bf Fig. 1}\ ]
One loop diagrams contributing to $\Pi_{33}$: Tadpole (a), ghost loop
(b), gluon loop (c). Notice the different line shapes for Polyakov
loop variables and charged gluons used throughout.

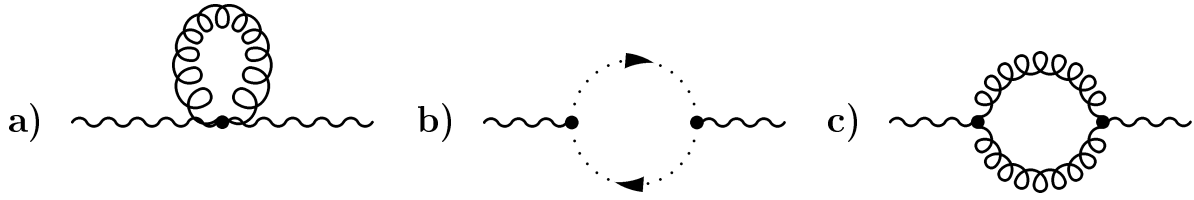
\item[{\bf Fig. 2}\ ]
A typical higher order diagram involving ghosts and Polyakov loop variables.

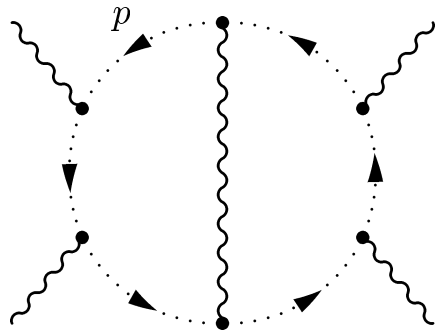
\item[{\bf Fig. 3}\ ]
One loop contribution to Polyakov loop correlator. The gluon-Polyakov loop
vertices are the shaded blobs.

\item[{\bf Fig. 4}\ ]
Gluon two point function to one loop. This and the following diagrams
pertain to the effective theory after integrating out the Polyakov loops.
Gluon loop (a) and tadpole (b).

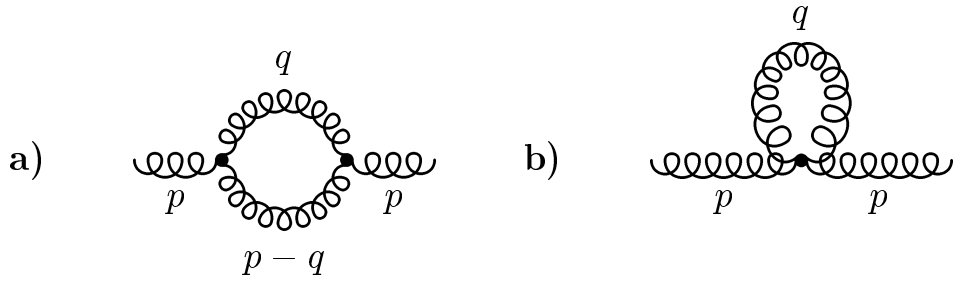
\item[{\bf Fig. 5}\ ]
Two loop Feynman diagrams contributing to Polyakov loop correlator.

\item[{\bf Fig. 6}\ ]
Contributions obtained from diagrams 5a, 5b by keeping the gauge terms
in the propagators everywhere except where the gluon line is marked.
Here, $g_{\mu \nu}$ term appears in one propagator only.

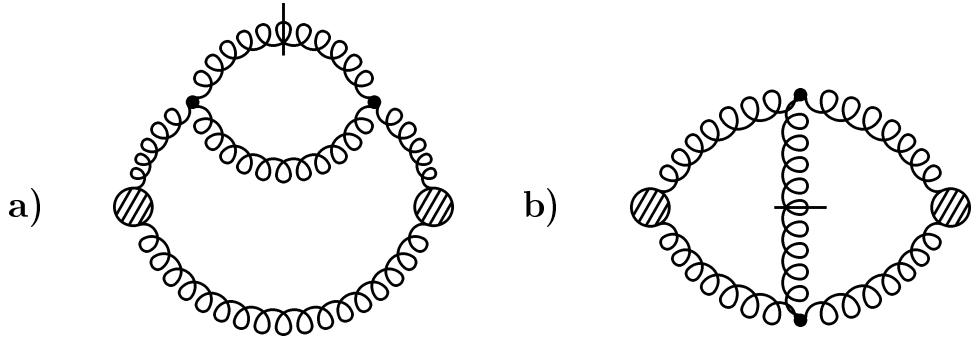
\item[{\bf Fig. 7}\ ]
Like Fig.~6a, but keeping the $g_{\mu \nu}$ term in two gluon propagators
in all possible ways in the two loop self energy diagram.

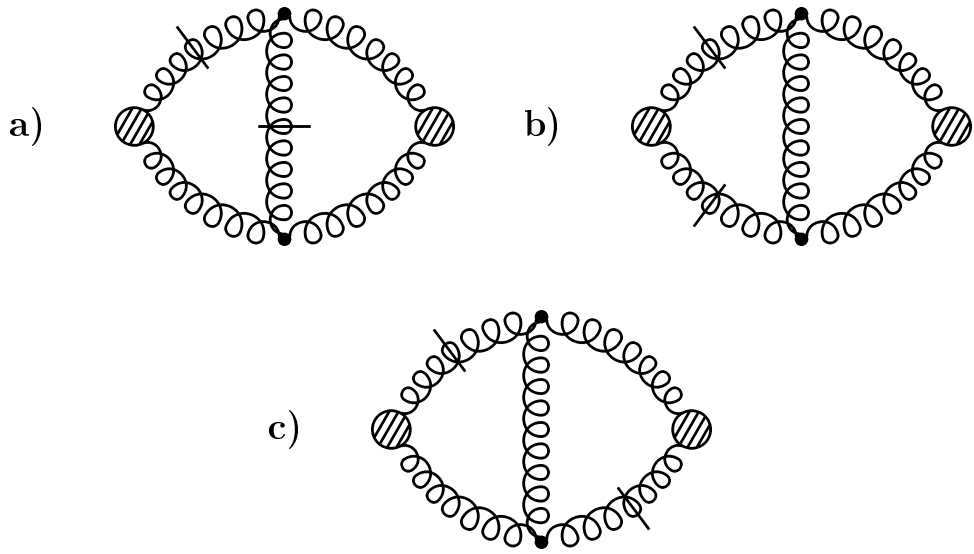
\item[{\bf Fig. 8}\ ]
Like Fig.~7, but for the two loop vertex correction diagram.

\end{itemize}

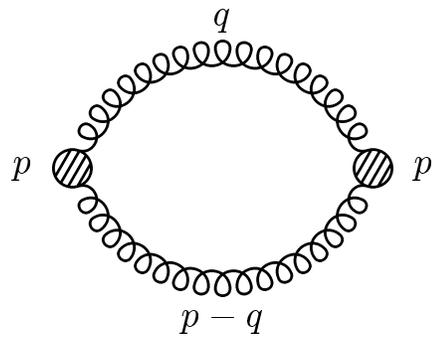
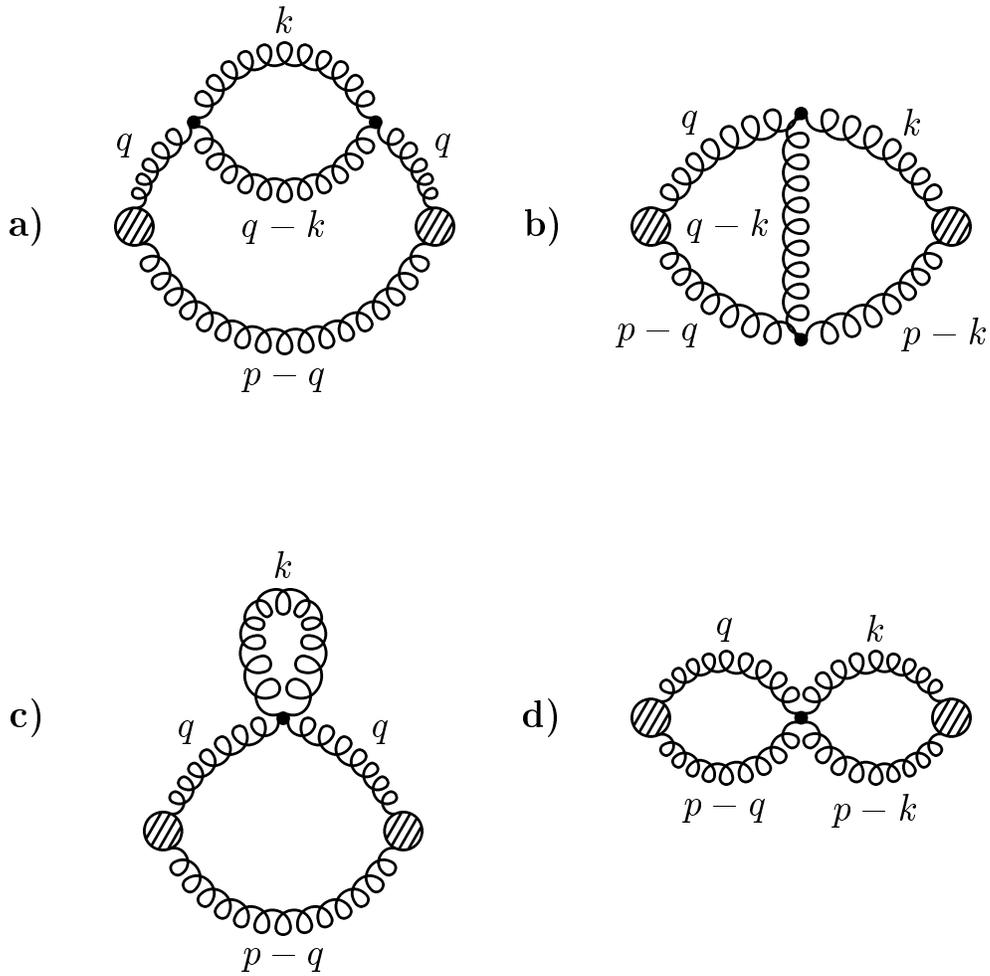
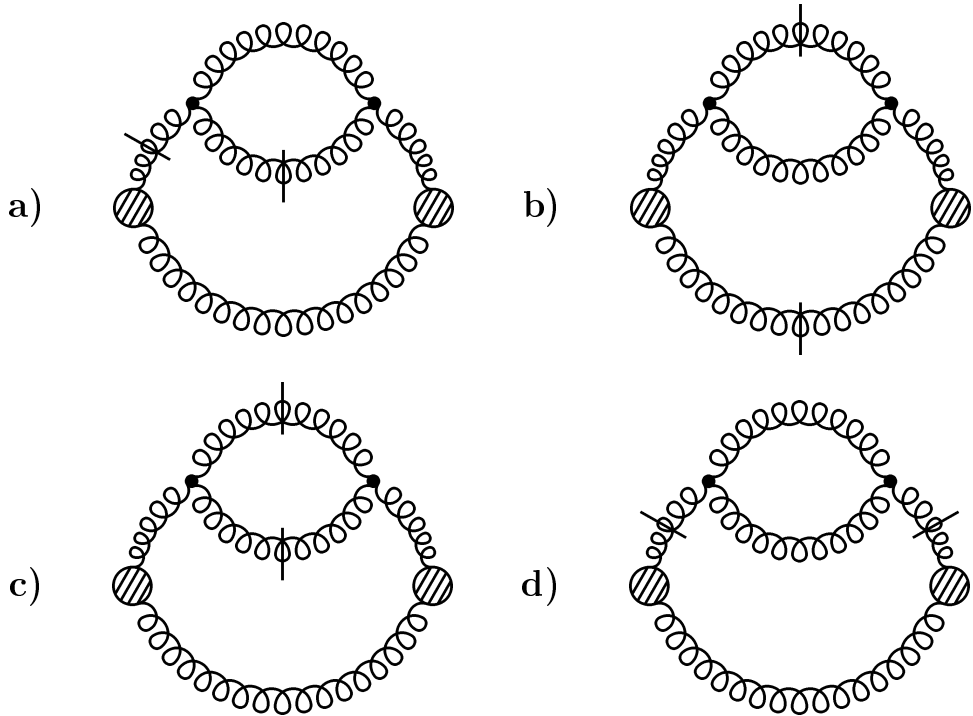
\end{document}